\documentclass[showpacs,aps,prb,reprint,superscriptaddress]{revtex4-2}
\usepackage{bm}
\usepackage{graphicx}
\usepackage{amsmath}
\usepackage{amsfonts}
\usepackage{amssymb}
\usepackage{mathtools}
\usepackage{epstopdf}
\usepackage{hyperref}
\usepackage{braket}
\usepackage{float}
\hypersetup{
    colorlinks,%
    citecolor=blue,%
    linkcolor=blue,%
    urlcolor=blue
}
\usepackage{tikz}\usetikzlibrary{petri}

\begin{document}

    \newcommand{\be}   {\begin{equation}}
    \newcommand{\ee}   {\end{equation}}
    \newcommand{\ba}   {\begin{eqnarray}}
    \newcommand{\ea}   {\end{eqnarray}}
    \newcommand{\ve}  {\varepsilon}

    \title{ Edge $\mathbb{Z}_3$ parafermions in fermionic lattices.}
        
    \author{Raphael L.~R.~C. Teixeira}
    \affiliation{Instituto de F\'{\i}sica, Universidade de S\~{a}o Paulo,
        C.P.\ 66318, 05315--970 S\~{a}o Paulo, SP, Brazil}
    \author{Luis G.~G.~V. Dias da Silva}
    \affiliation{Instituto de F\'{\i}sica, Universidade de S\~{a}o Paulo,
        C.P.\ 66318, 05315--970 S\~{a}o Paulo, SP, Brazil}
    
    \date{ \today }
    
    \begin{abstract}
    
    Parafermion modes are non-Abelian anyons which were introduced as $\mathbb{Z}_N$ generalizations of $\mathbb{Z}_2$ Majorana states. 
    In particular, $\mathbb{Z}_3$ parafermions can be used to produce Fibonacci anyons, laying a path towards universal topological quantum computation. 
    Due to their fractional nature, much of the theoretical work on $\mathbb{Z}_3$ parafermions has relied on bosonization methods or parafermionic quasi-particles. In this paper, we introduce a representation of $\mathbb{Z}_3$ parafermions in terms of purely fermionic models. We establish the equivalency of a family of lattice fermionic models written in the basis of the $t\!-\!J$ model with a Kitaev-like chain supporting free $\mathbb{Z}_3$ parafermionic modes at its ends. By using density matrix renormalization group calculations, we are able to characterize the topological phase transition and study the effect of local operators (doping and magnetic fields) on the spatial localization of the parafermionic modes and their stability. Moreover, we discuss the necessary ingredients towards realizing $\mathbb{Z}_3$ parafermions in strongly interacting electronic systems.

    \end{abstract} 
    \maketitle

    \section{Introduction}
    \label{sec:Intro}

Quasiparticle excitations with non-Abelian exchange statistics have long been proposed as the building blocks of topological quantum computers \cite{Nayak:NonAbelianAnyonTQC:2008}. Currently, the most promising  candidate to this role are Majorana bound states (MBSs) \cite{Kitaev:P.U:2001,Alicea:MilestoneMajoranaQC:2016}, with several proposals for their experimental realization being pursued \cite{Lutchyn:MajoranaReview:2018,Flensberg:ReviewMajorana:2021}. The basic mechanism is to use ``braiding operations'' of isolated MBSs to encode and process information in ``qbits" in a particular system. Such information is naturally protected by the non-Abelian nature of the braiding operations, which amounts to a ``topological protection''. 

A major limitation is that  not all unitary operations (``quantum gates'') needed for a fully functional quantum computer can be emulated by MBS braiding. As such, some of the operations would not be topologically protected (and thus more susceptible to noise and decoherence effects), hindering the prospect of an all-Majorana-based universal quantum computer \cite{Nayak:NonAbelianAnyonTQC:2008,Stern:TQC:2013,Nayak:MajoranaQuantumComputation:2015}. A possible solution to this problem is to use \emph{parafermionic bound states}, which are natural generalizations of MBSs and could in principle  be used as the building blocks of a universal topological quantum computer \cite{Stoudenmire:AssemblingFibonacci:2015,Loss:QCwithParafermion:2016,Hassler:ChangingAnyonicDegen:2016}.

The concept of parafermions was introduced by Fradkin and Kadanoff in the context of clock models in the early 1980's \cite{Fradkin:Parafermions:1980}. Parafermionic operators obey fractional statistics and also play an important role in critical behavior of clock models and $\mathbb{Z}_N$-symmetric models \cite{Fendley:ParafermionicZeroMode:2012}. Although parafermions were introduced in the context of statistical mechanics, they also arise as quasi-particle excitations in strong-correlated condensed matter \cite{Fleckenstein:ParafermionHEdgeQSHI:2018,Alavirad:Z3ParafermionWithoutAndreevFQH:2017,Vaezi:SuperconductingPFQH:2014,vonOppen:Z4parafermionQSHJosephsonImp:2017,Klinovaja:InteractingNanowire:2014}. 

In 1999, Read and Rezayi  showed that the excitations of some fractional quantum Hall states can be described using parafermions \cite{Read:ParafermionIncompressibleStates:1999}. Parafermions in $\mathbb{Z}_N$ models can have zero-energy edge states which are readily seen as a generalization of Majorana zero modes \cite{Fendley:ParafermionicZeroMode:2012,Fendley:StabilityParafermion:2014}. Although a topological computer made only by parafermions is not universal \cite{Mong:UniversalTQC:2014}, $\mathbb{Z}_3$ parafermions can be used to produce  ``Fibonacci anyons'' in two dimensions   \cite{Clarke:Non-AbelianAyonsFQH:2013,Stoudenmire:AssemblingFibonacci:2015}, leading to the interesting prospect of universal topological quantum computation using non-Abelian anyons \cite{Alicea:TopologicalPhasesWithParafermions:2016}. 

Over the years, several models that host isolated parafermionic modes in condensed matter systems have been considered. Proposals range from interacting quantum wires \cite{Klinovaja:TRIParafermionRashba:2014,Hughes:ParafermionicWiresCTstates:2017} to Abelian quantum Hall/superconductor hybrids and strongly interacting two-dimensional topological insulator edges coupled to superconductors \cite{Alicea:TopologicalPhasesWithParafermions:2016,Klinovaja:ParafermionsFTI:2014}. Moreover, several recent works focus on the capability of a given system to host zero-mode parafermions through low-energy models \cite{Klinovaja:InteractingNanowire:2014,Kane:Poorman:2015,Fleckenstein:ParafermionHEdgeQSHI:2018} and parafermion-only based Hamiltonians \cite{Mazza:AnyonicTBofparafermions:2019,Mazza:NonTopological:2018,Alicea:FermionezParafermionsSEMajorana:2018,Bernevig:ParafermionPhasesSbreakingTO:2016}. While the usual $\mathbb{Z}_2$ MBSs and, more recently, $\mathbb{Z}_4$ parafermionic zero modes \cite{Calzona:Z4FermionLattice:2018} emerge as quasiparticle excitations in purely fermionic models (and thus can be detected by their ``leakage'' to side-coupled electronic quantum dots \cite{Vernek:MajoarnaLeakage:2014,David:InteractionsMajoranaQD:2015,Teixeira:QD:2021}), the picture for  $\mathbb{Z}_3$ parafermionic modes is not quite clear. In fact, the question of whether purely fermionic models can produce (odd-$N$) $\mathbb{Z}_N$ parafermionic edge modes is still an open one.

In this paper, we address this fundamental question and propose a fermionic Hamiltonian that hosts $\mathbb{Z}_3$ parafermions. Our proposal is based on  strong local electron-electron interactions such that fermionic double occupancy  is suppressed. In a sense, the restriction to zero and singly occupied states plays a similar role as the restriction to ``spinless'' fermions  in the early proposals for Majorana bound states in semiconductor nanowires  \cite{Kitaev:P.U:2001,vonOppen:HelicalLiquidsMajoranaQW:2010,DasSarma:MajoranaPhaseSemiconductor:2010,Alicea:Reports:2012}.

In order to characterize the parafermionic phase, we use the density matrix renormalization group (DMRG) \cite{Schollwock:DMRG:2005,Schollwock:DMRG-MPS:2011,ITensor} to numerically calculate the energy gap, the entanglement entropy (EE), and the local spectral functions of the strongly correlated fermionic models. In addition, we study the stability of $\mathbb{Z}_3$ parafermions under effect of local doping and Zeeman terms, which can be important for the prospects of parafermion-based topological quantum computation. We find that the parafermionic zero modes are stable against such local perturbations, as long as these terms preserve the $\mathbb{Z}_3$ symmetry.

 The text is organized as follows. In Sec.~\ref{sec:modelmethods}, we introduce a strongly-correlated fermionic model displaying  $\mathbb{Z}_3$ parafermionic zero modes. By using a mean-field  {argument}, we derive an effective model which maps exactly into a Kitaev-like $\mathbb{Z}_3$ parafermion chain. Moreover, in  Sec.~\ref{sec:Equivalence} we show that the derived Hamiltonians indeed display a $\mathbb{Z}_3$ phase and can be deformed into each other without closing the gap. The effect of local operators is studied in Sec.~\ref{sec:LocalOp}, where we show the existence of a topological phase transition between a topological phase with edge parafermionic modes and a non-topological normal state. We also discuss the dependence of edge modes and their robustness under effects of local operators. To this end, we introduce a Fock-parafermion spectral function (FPF-SF) that can be used to distinguish the edge states.  Finally, we present our concluding remarks in Sec.\ \ref{sec:Conclusions} .  
 
 \section{Models and Methods}
 \label{sec:modelmethods}

  In this section, we propose a spinful fermionic model supporting $\mathbb{Z}_3$ parafermionic edge modes. As previously discussed, the intrinsic difficulty in devising such a system is that the Hamiltonian must conserve parity symmetry ($Z_2$) and, at the same time, conserve a $\mathbb{Z}_3$ symmetry of the parafermionic modes, two mutually exclusive requirements. A possible solution is that the $\mathbb{Z}_3$ parafermionic modes emerge in a situation in which the parity of the different ground states is set by the number of sites in the chain. As we shall see, this is indeed the case in the proposed fermionic lattice models.

 We begin by considering a fermionic spinful model with infinitely large  on-site Hubbard repulsive interactions. In this limit, we can safely exclude the doubly-occupied state in the local Hilbert space of each site, a procedure akin to that used in the derivation of the $t\!-\!J$ model \cite{Batista:GeneralizedJW:2001}). The Hamiltonian reads 
 \begin{equation}
 \label{eq:H246}
     H_{I}  =  H^{(2)} + H^{(4)} +  H^{(6)} \; ,
 \end{equation}
where
 \begin{equation}
 \label{eq:H(2)}
 H^{(2)} = \sum_{\substack{j=1\\\sigma={\uparrow,\downarrow}}}^{L-1} -t c_{\sigma,j}^\dagger c_{\sigma,j+1} -\Delta c^\dagger_{\sigma,j}c^\dagger_{-\sigma,j+1}+\mbox{ H.c.},
 \end{equation} 
 \begin{equation}
	\label{eq:H(4)}
	H^{(4)} =  -W_4 \sum_ {j=1}^{L-1} s^{+}_{j} s^{-}_{j+1}+\mbox{ H.c.} \;,
\end{equation}
 \begin{equation}
	\label{eq:H(6)}
	H^{(6)}= -W_6 \sum_ {j=2}^{L-1} s^{+}_{j-1} s^{+}_{j} s^{+}_{j+1}+\mbox{ H.c.}  \; .
\end{equation}

In the above, $s^{+}_{j} (s^{-}_{j})=c^{\dagger}_{\uparrow,j}c_{\downarrow,j} (c^{\dagger}_{\downarrow,j}c_{\uparrow,j})$ is the spin-flip operator, $t$ is the single-particle hopping, $\Delta$ is a $p$-wave-like superconducting order parameter that mixes spins in neighbor sites, $W_4$ is the strength of a synchronized spin-flip in two neighbor sites, while $W_6$ is the strength of synchronized spin flip in the three closest sites. We note that the  three-body interaction contained in $H^{(6)}$ is an important ingredient for the existence of $\mathbb{Z}_3$ parafermions.

The Hamiltonian $H_{I}$ given by Eq.~\eqref{eq:H246} has $S_3=\mathbb{Z}_3\rtimes \mathbb{Z}_2$ symmetry, where the $\mathbb{Z}_2$ part comes from spin flip and the $\mathbb{Z}_3$ component stems from the generalized three-valued ``parity'' operator:  

\begin{equation}
 \label{eq:PZ3}
\hat{P}_{\mathbb{Z}_3} = \omega^{\sum_{j=1}^{L}(n_{\uparrow,j}+2 n_{\downarrow,j})}
 \end{equation}
 where $\omega = e^{2 \pi i/3}$ and $n_{\sigma,j} = c^\dagger_{\sigma,j} c_{\sigma,j}$ is the usual fermionic number operator at site $j$. 
 
 One can readily check that $\hat{P}^{\dagger}_{\mathbb{Z}_3} H_{I} \hat{P}_{\mathbb{Z}_3} = H_{I}$. {More importantly, as further discussed in Sections \ref{sec:Equivalence} and \ref{sec:LocalOp}, the ground states of $H_{I}$ are also eigenstates of $\hat{P}_{\mathbb{Z}_3}$ and, in the $\mathbb{Z}_3$ phase, two out of the three ground states are related by a  spin-flip transformation. These states can be split by an out-of-plane local Zeeman term which breaks the corresponding $\mathbb{Z}_2$ symmetry.}

{ The next step is to check under which conditions $H_{I}$ can be related to } a benchmark Hamiltonian supporting $\mathbb{Z}_3$ parafermionic edge modes.  {To this end, } we can use a Kitaev-like parafermion chain~\cite{Fendley:ParafermionicZeroMode:2012,Zhuang:PhaseDiagramZ3Parafermion:2015}, the Hamiltonian of which is given by
 \begin{equation}\label{eq:pf}
 H_{pf}=-J\sum_{j=1}^{L-1} \psi_j\chi^\dagger_{j+1} +\mbox{H.c.},
 \end{equation} 
where each site has two parafermion modes $\psi$ and $\chi$ satisfying parafermionic identities $\psi_j^\dagger = \psi_j^2$, $\chi_j^\dagger = \chi_j^2$ and $\chi_j \psi_j = \omega\psi_j \chi_j$. For different sites, they satisfy a parafermionic exchange algebra $\psi_l \psi_j = \omega \psi_j \psi_l$, $\chi_l \chi_j = \omega \chi_j \chi_l$  and $\chi_l \psi_j = \omega \psi_j \chi_l$ for $l< j$. 

This model is exactly solvable for any $J>0$, showing a threefold ($\mathbb{Z}_3$ symmetric) degenerate ground state \cite{Fendley:ParafermionicZeroMode:2012,Zhuang:PhaseDiagramZ3Parafermion:2015,Mazza:ParafermionGS:2017}. Moreover, one can show that $H_{pf}$ can be written in terms of fermionic operators (see Appendix~\ref{Appendix:Fermionization}) yielding a similar Hamiltonian as that of Eq.~\eqref{eq:H246}.

We note that fermionization of $H_{pf}$ produces a parity-violating  interaction term $H^{(3)}$ given by  
 \begin{align}
 \label{eq:H(3)}
 H^{(3)}=& -W_3 \sum_ {\substack{j=1\\\sigma={\uparrow,\downarrow}}}^{L-1} (-1)^{\sum_{p<j}n_p} \Big [c_{\sigma,j}c^\dagger_{-\sigma,j+1}c_{\sigma,j+1} +\\& +c^\dagger_{\sigma,j}c_{-\sigma,j}c^\dagger_{\sigma,j+1}\Big]+ H.c.\nonumber \; .
 \end{align}

This term corresponds to an exotic process of creation (annihilation) of an electron together with a spin flip in the neighboring site. As such, it does not conserve either parity or electron number. In fact, this term can be understood as an {approximation of the} mean-field interaction of the term $H^{(6)}$ given by Eq.~\eqref{eq:H(6)} in which the parity is spontaneously broken (see Appendix~\ref{Appendix:MF}) .

We thus define Hamiltonian $H_{II}$ as 
\begin{equation}
 \label{eq:H243}
     H_{II}  =  H^{(2)} + H^{(4)} +  H^{(3)} \; .
 \end{equation}

For $J \! :=\! t \!=\! \Delta \!=\! W_4 \!=\! W_3$,  it can be shown that $H_{II} \! \rightarrow \! H_{pf}$ (see Appendix~\ref{Appendix:Fermionization} for details). {In this special limit, the ground state can be obtained analytically~\cite{Mazza:ParafermionGS:2017}. }

We should point out that the long-range interaction terms in $H_{II}$ do not prevent the existence of a topological phase \cite{Gong:Phys.Rev.B:041102:2016,Yu:Phys.Rev.B:245131:2020}. In the present case, not only are there free parafermionic edge operators ($\chi_1$ and $\psi_L$) which couple the different ground states but also the ground states are indistinguishable by local probes, satisfying the criteria for topological order \cite{Bernevig:ParafermionPhasesSbreakingTO:2016}. 

{Although $H_{II}$ can be obtained from a mean-field-like form of $H_{I}$, it is not \emph{a priori} clear that $H_{I}$ should have a $\mathbb{Z}_3$ topological phase. The deep connection of the fermionic model of Eq.~\eqref{eq:H246} and the parafermionic chain of Eq.~\eqref{eq:pf} constitutes one of the main results of this paper and it is discussed in detail in Section~\ref{sec:Equivalence}.}

  \section{Equivalence of the models}
 \label{sec:Equivalence}

  The existence of a $\mathbb{Z}_3$ phase in the Hamiltonian $H_I$ can be established by two complementary methods. First, we show that there is a phase transition in which the ground state of the system goes from nondegenerate to threefold degenerate. Then we show that in this threefold degenerate phase it is possible to smoothly deform $H_{I}$ into $H_{II}$ in a regime of parameters in which it displays the same $\mathbb{Z}_3$ parafermion phase as  $H_{pf}$. 
  
{These two methods, together with the existence of gapless edge states and the indistinguishability of ground states by local operators (discussed in Sec.~\ref{sec:FPFSpecFun}) and the protection against disorder and single impurities that preserve $\mathbb{Z}_3$ symmetry (Sec.~\ref{sec:LocalOp}) are strong evidences of the existence of a $\mathbb{Z}_3$  topological phase in both $H_I$ and $H_{II}$}. 

 \subsection{Gap closing at the transition}
 \label{sec:gapclosing}
 
 The different phases can be characterized by two main quantities: the ground-state degeneracy  $n_{\rm gs}$ and the energy gap $E_{\rm gap}$  between the ground state and the first excited (many-body) state. To this end, the  ground states of the fermionic Hamiltonians are calculated for the different model parameters with the DMRG method~\cite{White:Phys.Rev.B:10345:1993,Schollwock:DMRG:2005,Schollwock:DMRG-MPS:2011} via the \textsc{itensor} package \cite{ITensor}.  In the remainder of the paper, we use $t\!=\!\Delta\!=\!W_4$ and a 100-site chain, unless otherwise specified.

 We obtain $n_{\rm gs}$ in the DMRG calculations by counting the number of low-lying states within a window $\delta E \lesssim 10^{-3} t$ of the ground-state energy (the hopping $t$ is the energy unit). This value is well within the ground-state energy  accuracy in the DMRG calculations given the bond dimension and system's size. It is also enough to calculate $E_{\rm gap}$: for the parameters used, $E_{\rm gap} \gtrsim 10^{-2}t$ for all cases.

   \begin{figure}[t]
 	\begin{center}
 		\includegraphics[width=1\columnwidth]{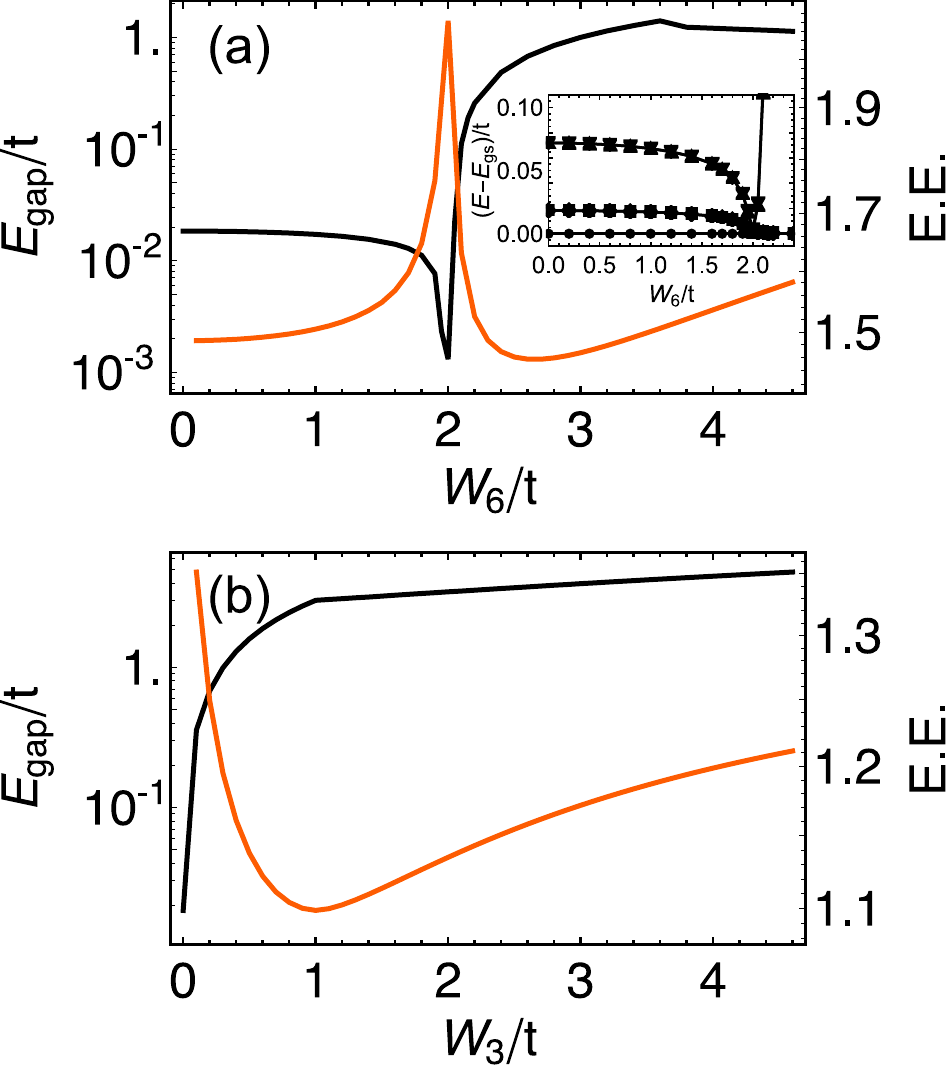}
 		\caption{Gap (black) and entanglement entropy (red) as a function of interaction strength (a) $W_6$ and (b) $W_3$ for the models described by $H_I$ and $H_{II}$ respectively. While the phase transition occurs at $W_6>2t$ in $H_{I}$, any $W_3>0$ induces the $\mathbb{Z}_3$ phase in $H_{II}$. {The inset in panel (a) shows the difference in energy level between the 5 states with lower energy ($E$) and the ground state energt $E_{\rm gs}$. Note that all the states converge around the phase transition.} } 
 		\label{fig:phases}
 	\end{center}
 \end{figure}

Results for $E_{\rm gap}$ for Hamiltonians $H_{I}$ and $H_{II}$ as a function of $W_6$ and $W_3$ respectively are shown in Fig.~\ref{fig:phases}. 
Figure \ref{fig:phases}(a) shows $E_{\rm gap}$ for $H_{I}$ as a function of $W_6$. For small values of $W_6$, the system is in the trivial regime and has a small gap ($\sim 10^{-2}t$) which we attribute to finite-size effects. For $W_6\!\approx\!2t$ \footnote{Around $W_6\!=\!2t$ we used steps of $0.05t$ to plot figure \ref{fig:phases}(a).}, the system undergoes a phase transition, characterized by a sharp decrease in $E_{\rm gap}$ (``gap closing").

{The phase transition becomes evident by plotting the low-lying energy levels as a function of $W_6$ (inset Fig.~\ref{fig:phases}(a)). For  $W_6 \gtrsim 2t$, the single ground state and a pair of higher energy states merge, abruptly increasing the degeneracy from $n_{\rm gs} \! = \! 1$  to $3$. The other higher energy states also close the gap at the same point and reopen, indicating the bulk gap closes at the transition.} 

The threefold ground-state degeneracy characterizes the new phase as a  $\mathbb{Z}_3$ (topological) phase. The  {closure of the gap} shown in Fig.~\ref{fig:phases}(a) is accompanied by a discontinuity in the first derivative of the entanglement entropy EE~\footnote{To compute the entanglement entropy, we perform DMRG calculations fully preserving the $\mathbb{Z}_3$ symmetry. The entanglement entropy is then calculated at the system's center bond.}.

 At the $\mathbb{Z}_3$ phase, all ground states have a well defined $\mathbb{Z}_2$  parity which depends on the length $L$ of the chain as $\hat{P}_{\mathbb{Z}_2}\!=\!(-1)^{\sum_i^L (n_{\uparrow,i}+n_{\downarrow,i})}\!=\!(-1)^{\text{L mod} 2}$. Moreover, these ground states are characterized by a $\mathbb{Z}_3$ parity operator $\hat{P}_{\mathbb{Z}_3}\!=\!\omega^{\sum_i^L(n_{\uparrow,i}+2 n_{\downarrow,i})}$ defined in Eq.~\eqref{eq:PZ3}.  The different ground states can be distinguished by the respective eigenvalue of $\hat{P}_{\mathbb{Z}_3}$, which can be $1,\omega$ or $\omega^2$. For this reason, the $\mathbb{Z}_3$ phase can not be understood as a simple combination of a Majorana-hosting phase together with a $\mathbb{Z}_2$ broken symmetry phase as it is the case for $\mathbb{Z}_4$ parafermions~\cite{Calzona:Z4FermionLattice:2018,Alicea:FermionezParafermionsSEMajorana:2018,Teixeira:QD:2021}. An important consequence  is that a spin-flip transformation swaps the sectors $\langle \hat{P}_{\mathbb{Z}_3}\rangle\!=\!\omega$ and $\omega^2$, while $\langle \hat{P}_{\mathbb{Z}_3}\rangle\!=\!1$ is mapped into itself. This translates into the formation of doublets in the excited states.

A similar analysis can be made for $H_{II}$, by plotting $E_{\rm gap}$ for increasing $W_3$ with $t=\Delta=W_4$ (Fig.~\ref{fig:phases}(b)). The main difference is that the critical value in which the system goes from the trivial to $\mathbb{Z}_3$ phase is $W_3\!=\!0$. As such, for any $W_3>0$ the system is in the $\mathbb{Z}_3$ phase and no phase transition takes place for nonzero values of $W_3$, as indicated by the absence of a peak in the entanglement entropy. In fact, the EE reaches its minimum value (log(3)) for $W_3\!=\!t$, precisely the point where the mapping of $H_{II}$ to the parafermion chain Hamiltonian $H_{\rm pf}$ is exact. 

 \subsection{Entanglement spectrum and finite-size effects }
 \label{sec:entanglementspectrum}

    \begin{figure}[t]
 	\begin{center}
 		\includegraphics[width=0.8\columnwidth]{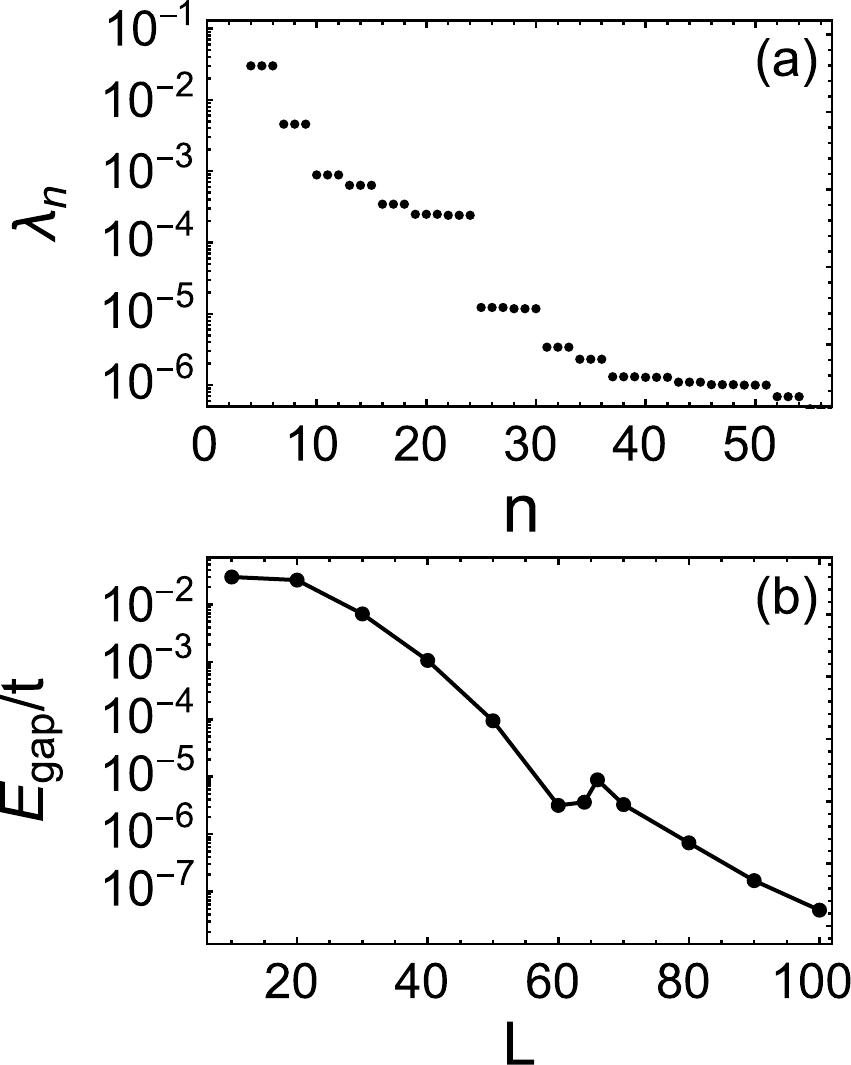}
 		\caption{Characterization of $H_I$ with $W_6=3.2t$. (a) Entanglement spectrum for the ground state in the sector $\langle \hat{P}_{\mathbb{Z}_3}\rangle\!=\!1$ for a chain with $L=100$ sites; the spectrum has a threefold degeneracy and is the same for other sectors. (b) Gap between the ground states due to the finite size of a chain with $L$ sites; note that there are two different behaviors for $L\lesssim60$ and $L\gtrsim70$ indicating two mechanisms of intraedge interaction.} 
 		\label{fig:EntangSpec}
 	\end{center}
 \end{figure}

{In order to better understand the nature of the ground state, we turn to the entanglement spectrum of $H_{I}$. Fig.~\ref{fig:EntangSpec}(a) shows the largest eigenvalues of the reduced density matrix ($\lambda_n > 5\!\times\!10^{-6}$) calculated with DMRG in the sector $\langle \hat{P}_{\mathbb{Z}_3}\rangle\!=\!1$  for $W_6\!=\!3.2t$ and $L\!=\!100$ sites.  As a general feature, the spectrum is threefold degenerate, as it would be expected for a $\mathbb{Z}_3$ phase \cite{Mazza:NonTopological:2018}. The same result is obtained for the other parity sectors. }

{ We point out in passing that a similar result is obtained for $H_{II}$. In particular, at the point $W_3\!=\!t$, the system is maximally entangled with  bond dimension 3, i.e., only three nonzero $\lambda_n$, all equal to $1/3$.}

{Due to intraedge coupling of the edge states, the gap calculation of $H_{I}$ is prone to finite-size effects, as shown in Fig.~\ref{fig:EntangSpec}(b). We note that there are two distinct regimes, with the gap decaying faster for $L \gtrsim 70$ sites. This indicates that the decay for $L \lesssim 60$ is mainly due to a decreasing in the intraedge overlap of the edge modes, which is also consistent with the spectral function results discussed in Section \ref{sec:FPFSpecFun}. }

{This is in striking contrast with the case of $H_{II}$, where the edge modes are much more localized. In fact, since $H_{II}$ is exactly mapped in $H_{pf}$, there is no dependence of the gap size with the chain length. This is clearly not the case for $H_{I}$, which needs large chains such that finite-size effects can be neglected.}

 \subsection{Deforming $H_{I}$ into $H_{II}$ }
 \label{sec:deformation}

In order to confirm that the limits of large $W_6$ for $H_{I}$ and large $W_3$ for $H_{II}$ correspond to the same topological $\mathbb{Z}_3$ phase, we consider the following Hamiltonian:
 \begin{equation}\label{eq:deformation}
 H^{\prime}(x)= (1-x) H_{I} + x H_{II},
 \end{equation}
 which is equal to $H_{I}$ and $H_{II}$ for $x\!=\!0$ and $1$ respectively. In a sense, $x$ acts as a parameter which continuously ``deforms" $H^{(6)}$ into $H^{(3)}$.

 \begin{figure}[t]
 	\begin{center}
 		\includegraphics[width=1\columnwidth]{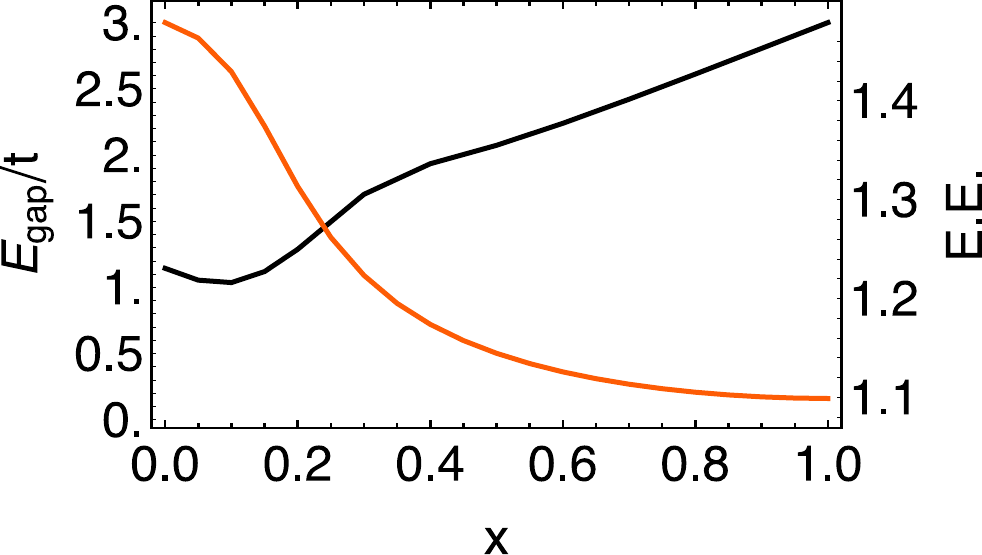}
 		\caption{Gap (black) and entanglement entropy (red) as a function of deformation parameter $x$, Eq.~\eqref{eq:deformation}. We consider the case $W_6=3.2t$ and $W_3=t$.} 
 		\label{fig:equi}
 	\end{center}
 \end{figure}

 Figure \ref{fig:equi} shows the dependency of $E_{\rm gap}$ and entanglement entropy with $x$ for $W_6=3.2t$ and $W_3=t$. The crucial result is that there is no gap closing or sharp features in the entanglement entropy (which would be indicatives of a phase transition) as $x$ varies from $0$ to $1$.  In fact, the minimum gap is $E_{gap}\approx t$ for $x=0.1$ and the difference in the entanglement entropy's value is due to the differences in the ground-state occupancies. This shows that  both Hamiltonians describe the same   $\mathbb{Z}_3$ topological phase for these values of $W_6$ and $W_3$.

 \section{Effects of local operators}
 \label{sec:LocalOp}

Although both $H_{I}$ and $H_{II}$ display a parafermion-hosting $\mathbb{Z}_3$ phase, the ground states themselves are very different. For example, the $H_{I}$ ground state has a well-defined parity, while the $H_{II}$ ground state does not. Nonetheless, we expect the general behavior of parafermions under changes in local operators to be similar. 

Since the DMRG calculations for $H_{II}$ run a few orders of magnitude faster as compared to $H_{I}$, we will use $H_{II}$ as a ``benchmark" in this Section. Unless otherwise stated, we set $W_3\!=\!t$ (and $W_6\!=\!0$), meaning that, in the absence of other terms in the Hamiltonian, the system will be in the topological phase of $H_{II}$.

Previous studies \cite{Fendley:ParafermionicZeroMode:2012,Zhuang:PhaseDiagramZ3Parafermion:2015}  have shown that local interactions might be able to destroy the parafermion phase. Specifically, the interaction $-f(e^{i\theta}\psi^\dagger_j\chi_j \!+\! e^{-i\theta}\chi^\dagger_j\psi_j)$ destroys the parafermion edge while conserving the FPF number. In particular, there is a phase transition at $f\!=\!J$  with $\theta\!=\!0$. This interaction translates into the fermionic language (see Appendix \ref{Appendix:Fermionization}) as
 \begin{align}
 e^{i\theta}\psi^\dagger_i \chi_i + e^{-i\theta}\chi^\dagger_i \psi_i = -&2\sqrt{3}\sin(\theta) \ n_{\uparrow,i}+\\ \nonumber
 &+[3cos(\theta)-\sqrt{3}\sin(\theta)]n_{\downarrow,i} \; ,
 \end{align}  
 which can be thought of as a mixing of magnetic field and chemical potential for any $\theta$. This shows the importance of local operators to parafermions. In particular, we are interested in the effects of chemical potential,
 \begin{equation}
 H_d = -\sum_{\substack{j=1\\\sigma={\uparrow,\downarrow}}}^L \mu n_{\sigma,j},\label{eq:doping}
 \end{equation}
 and Zeeman fields in all three directions: 
  \begin{align}
 H_x =& \sum_{\substack{j=1\\\sigma={\uparrow,\downarrow}}}^L V_x c^\dagger_{\sigma,j}c_{-\sigma,j} \;,\label{eq:Zimman-x}\\
 \nonumber\\
 H_y =& \sum_{\substack{j=1\\\sigma={\uparrow,\downarrow}}}^L - i V_y \sigma c^\dagger_{\sigma,j}c_{-\sigma,j} \; ,\label{eq:Zimman-y}\\
  \nonumber\\
 H_z =& \sum_{\substack{j=1\\\sigma={\uparrow,\downarrow}}}^L V_z \sigma n_{\sigma,j} \; , \label{eq:Zimman-z} 
 \end{align}
 which will be added to $H_{II}$.

\subsection{Gap closing \label{sec:Gap}}

\begin{figure}[t]
	\begin{center}
		\includegraphics[width=1\columnwidth]{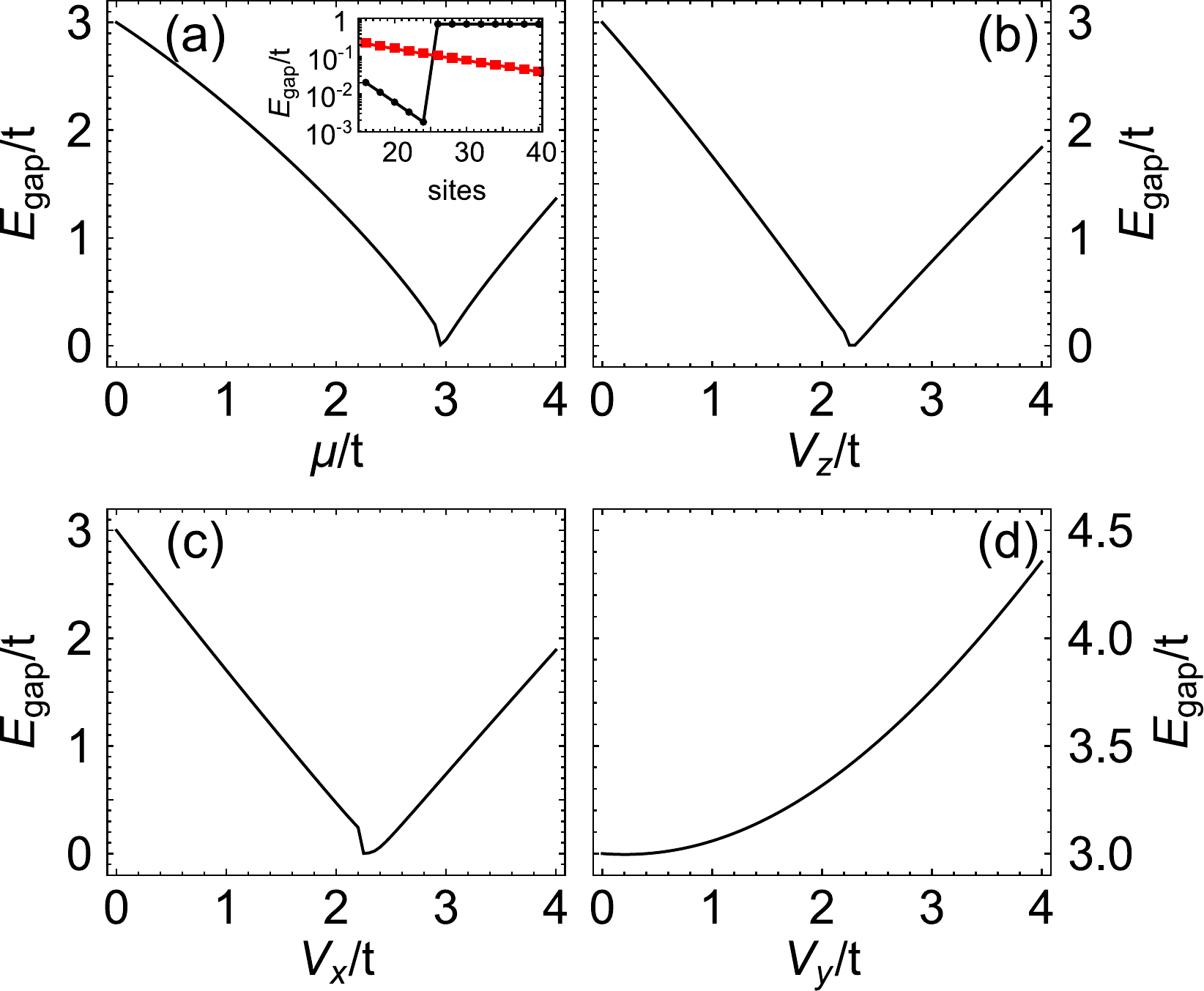}
		\caption{Dependence of the gap with respect to local operators. (a) Doping ($\mu$), (b) $z$-direction ($V_z$), (c) $x$-direction ($V_x$), and (d) $y$-direction ($V_y$) Zeeman terms. Panels (a) and (b) show a topological phase transition at $\mu=3t$ and $V_z=2.3t$, respectively. Panel (c) has a phase transition between a twofold degenerate state and a normal state for $V_x\approx2.5J$. Finite-size effects are responsible for the discontinuity in the gap. Panel (d) shows no phase transition as a function of $V_y$, and the gap always increases. The inset in panel (a) shows the exponential dependence of the gap between the parafermionic modes with the chain length for $\mu=2.5t$ (black circles) and $\mu=2.9t$ (red squares). } 
		\label{fig:gap}
	\end{center}
\end{figure}

Figure \ref{fig:gap} shows the dependence of the gap energy $E_{\rm gap}$  with each of these local terms for a 100-site chain. Regarding the chemical potential ($\mu$), we see a near gap closing at $\mu \! \approx \! 2.9t$ (Fig.~\ref{fig:gap}(a)). This is consistent with previous results \cite{Zhuang:PhaseDiagramZ3Parafermion:2015} which show a transition between topological (parafermionic) and normal phases for $\mu \!=\! 3t$. We believe the small discrepancy with our result can be accounted for by finite-size effects. In fact, we can verify that the phase transition point approaches $\mu \!=\! 3t$ for  increasing the chain sizes (see Section~\ref{sec:FPFSpecFun}).
 
The inset of Fig.~\ref{fig:gap}(a) shows that the gap between the parafermion states decreases exponentially in the topological phase.  This gap arises due to the coupling between parafermionic modes located at both ends of the chain. This exponential decay depends on $\mu$ \footnote{The exponential decay happens as long as $\mu$ is such that does not break the $\mathbb{Z}_3$ symmetry.}, as illustrated by the slower decay with size by $\mu \!=\! 2.9t$ as compared to the case $\mu \!=\! 2.5t$. As such, this is equivalent to the ``exponential protection" predicted for Majorana modes \cite{DasSarma:SmokingGun:2012}
and was also predicted to occur for  $\mathbb{Z}_4$ parafermions as well \cite{Calzona:Z4FermionLattice:2018}.

 Parafermionic edge modes are also stable under a small local Zeeman-like term in the $z$ direction proportional to $V_z$, as given by Eq.~\eqref{eq:Zimman-z}. As shown in Fig.~\ref{fig:gap}(b), the topological phase is destroyed only for relatively large values of the Zeeman term ($V_z  \gtrsim 2.3t$). In addition, similarly to the dependency with the doping $\mu$, the transition value is sensitive to {finite-size} effects even for long chains. In both cases ($\mu$ and $V_z$), the ground state goes from a threefold degenerate to a nondegenerate one at the transition.

As discussed in detail in Appendix~\ref{Appendix:Fermionization}, a generic magnetic field in the $xy$ plane dos \emph{not} conserve the Fock-parafermion number, thus breaking the $\mathbb{Z}_3$ symmetry. In fact, any small positive  Zeeman term in the $x$-direction $(V_x > 0)$ breaks the ground-state $\mathbb{Z}_3$ symmetry, changing the ground-state degeneracy from $n_{\rm gs}\!=\!3$ to $2$. As $V_x$ increases, a second phase transition occurs, further reducing $n_{\rm gs}$ from 2 to  1 . This is shown in Fig.~\ref{fig:gap}(c), where the phase transition to the $n_{\rm gs}\!=\!1$ (nondegenerate) ground state occurs around $V_x\approx2.5t$. In this case, finite-size effects are more prominent than the previous cases, making it difficult to pinpoint the exact  $V_x$ value where phase transition occurs for large systems.

By contrast, any positive Zeeman term in the $y$ direction ($V_y >0$) produces a phase transition directly from from $n_{\rm gs}\!=\!3$ to $1$. The gap increases monotonically with $V_y$, as shown in Fig.~\ref{fig:gap}(d). Due to these differences between $x$ and $y$ directions, we expect a strong dependence of the gap with the direction of magnetic fields in the $xy$ plane.

   \begin{figure}[t]
 	\begin{center}
 		\includegraphics[width=1\columnwidth]{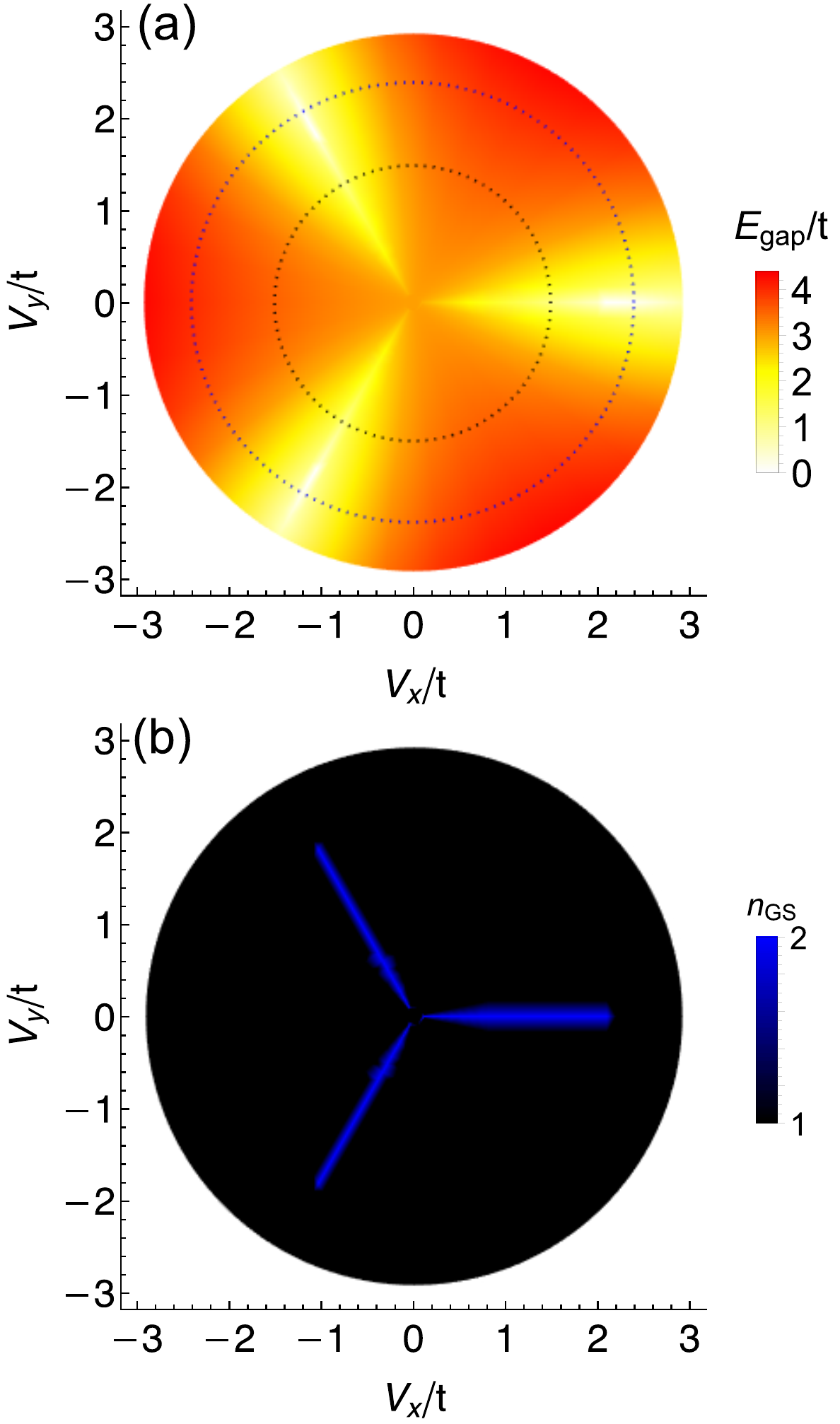}
 		\caption{Dependence of (a) the gap to the intensity and angle of the magnetic field and (b) its ground-state degeneracy, $n_gs$. For fixed $V_{xy}<2.5t$ the gap is minimum at angles $0,\pm2\pi/3$ and the ground state is twofold degenerate. The dotted circles in panel (a) correspond to the transversal cut shown in Fig.~\ref{fig:cutSxy}.} 
 		\label{fig:Sxy}
 	\end{center}
  \end{figure}

 In order to better understand how the local Zeeman terms in the $xy$ plane affect the parafermionic chain ground state, we consider a generic Zeeman term arising from a magnetic field in the $xy$ plane, given by
  \begin{gather}\label{eq:xyplane}
 H_{xy} = V_{xy}\sum_{j=1}^L e^{-i \theta}c^\dagger_{\uparrow,j}c_{\downarrow,j} + e^{i \theta}c^\dagger_{\downarrow,j}c_{\uparrow,j},
 \end{gather}
where $V_{xy}\!=\!\sqrt{V_x^2 \!+\!V_y^2}$ is the strength of the Zeeman field and $\theta$ is the magnetic field angle with respect to the $x$ direction. 
 
We calculate the gap energy ($E_{\rm gap}$) and ground-state degeneracy ($n_{\rm gs}$) as a function of $V_{xy}$ and $\theta$. The results are depicted in Fig~\ref{fig:Sxy}. Notice the clear symmetry in $E_{\rm gap}(\theta)$ and $n_{\rm gs}(\theta)$ as $\theta\to\theta+2\pi/3$. This is in fact due to the invariance of  $H_{xy}$ under a $2\pi/3$ rotation, associated with the $\mathbb{Z}_3$ symmetry of the full Hamiltonian. This invariance becomes clear by writing  Eq.~\eqref{eq:xyplane} in terms of parafermion operators, which can be accomplished by inverting  the fermionization process discussed in  Appendix~\ref{Appendix:Fermionization}. The result is
 \begin{align}
 H_{xy} = \frac{V_{xy}}{3}\sum_{j=1}^L \omega^{\sum_{p<j}N_p}e^{-i \theta}\left[\chi_j +\omega \psi_j + \chi_j^\dagger \psi_j^\dagger \right] \nonumber\\ + \omega^{2\sum_{p<j}N_p}e^{i \theta}\left[\chi_j^\dagger +\omega^2 \psi_j^{\dagger} + \omega^2 \chi_j \psi_j \right].
 \end{align}

 For $\theta \!=\! 0$ and $\pm2\pi/3$, the Hamiltonian is invariant under a transformation $\chi\to e^{i \theta}\chi$ and $\psi\to e^{i \theta}\psi$. This can be easily seen in Fig.~\ref{fig:Sxy}, where the smallest gaps occur at angles $\theta^{\rm min}_n = 2n\pi/3$, $n=1,2,3$. In those cases, the ground state is doubly degenerate, as discussed above, with a phase transition occurring at $V_{xy} \approx 2.5t$, similar to that shown in Fig.~\ref{fig:gap}(c).
 
  \begin{figure}[t]
 	\begin{center}
 		\includegraphics[width=1\columnwidth]{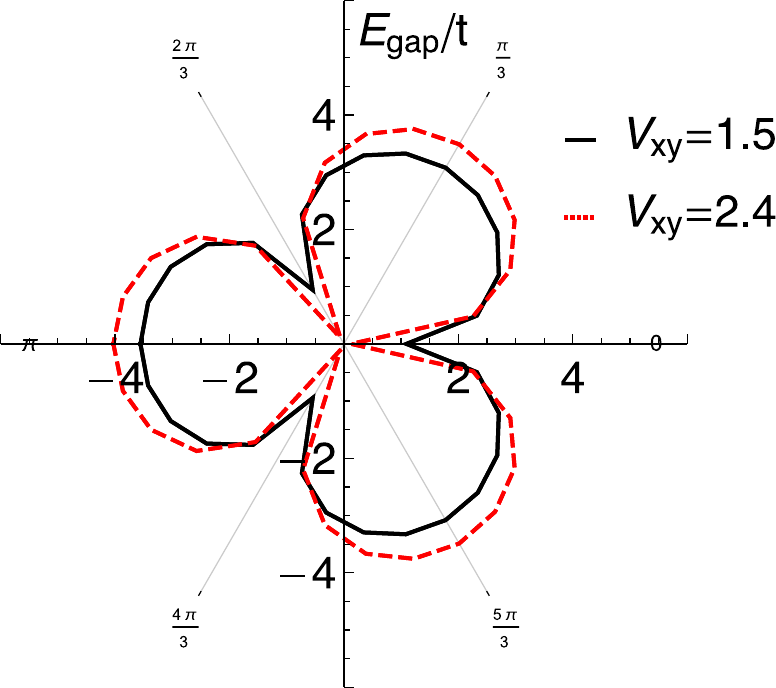}
 		\caption{Transversal cut of Fig.~\ref{fig:Sxy} with $V_{xy}=1.5t$, solid black, and $V_{xy}=2.4t$, dashed red.} 
 		\label{fig:cutSxy}
 	\end{center}
 \end{figure}

 In order to better visualize this, a crosscut of Fig.~\ref{fig:Sxy} with $V_{xy}=1.5t$ and $2.4t$ is shown in Fig.~\ref{fig:cutSxy}. The minimum gap occurs at $\theta^{\rm min}_n = 2n\pi/3$  and the maxima are at $\theta^{\rm max}_n = (2n-1)\pi/3$  with $n=1,2,3$.

\subsection{Local disorder \label{sec:Disorder}}

 We also considered the case of a locally disordered potential, i.e., the on-site terms $\mu$ or $V_z$ are randomly distributed. We simulated 20 different profiles of chemical potential or Zeeman in the $z$ direction; both $\mu_j$ and $V_{z,j}$ were generated from a uniform distribution and varying the mean values of $\mu/t$ or $V_z/t$.

\begin{figure}[t]
	\begin{center}
		\includegraphics[width=1\columnwidth]{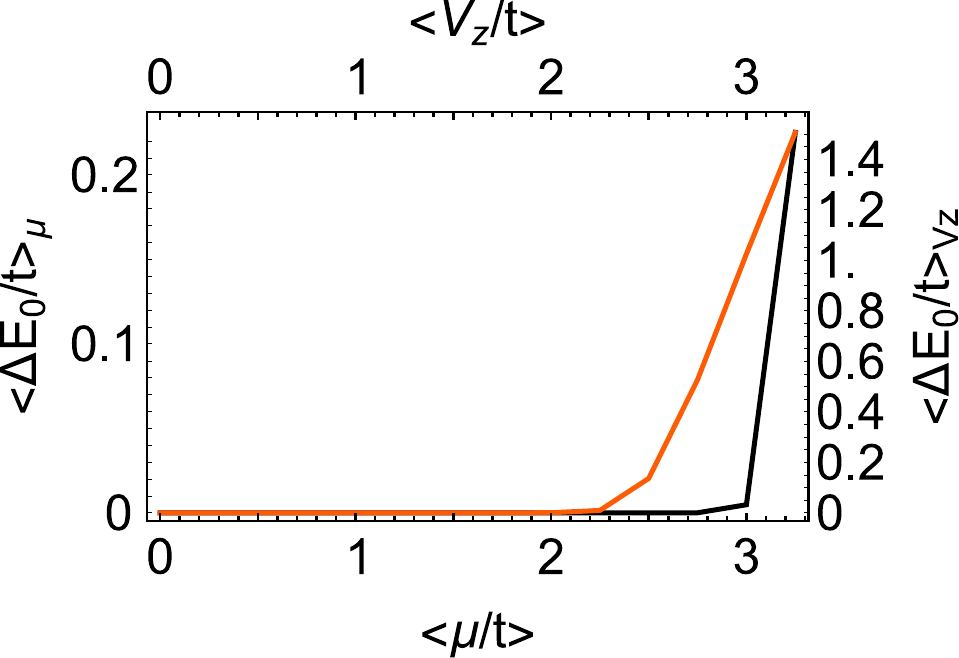}
		\caption{Mean energy splitting between the ground states due to random potential with maximum intensity  for the chemical potential $\mu$ (solid black) and Zeeman field at $z$ direction $V_z$ (dashed red). The mean was calculated based on 20 different distributions of impurities. Note the splitting happens for values of $\langle \mu \rangle/t$ or $\langle V_z \rangle/t$ of the order of the critical value seen in Fig.~\ref{fig:gap}.} 
		\label{fig:disorder}
	\end{center}
\end{figure}

 Figure \ref{fig:disorder} shows the variation of the mean disorder-induced splitting $\langle \Delta E_{0} \rangle$ between the ground states as a function of the mean values of $\mu$ or $V_z$. The splitting is zero (i.e., the parafermionic phase is \emph{not} destroyed) provided that the mean values are much smaller than the critical values for which the system undergoes a phase transition, shown in Fig.~\ref{fig:gap}. By contrast, for disordered chains with mean values of $\langle \mu \rangle/t$ or $\langle V_z \rangle/t$ close to critical values, even a handful of sites are enough to open a gap and lift the ground-state degeneracy.

Nonetheless, a single impurity symmetry in the bulk do not lift the threefold degeneracy as long as it preserve the $\mathbb{Z}_3$ symmetry. This property holds even when the impurity potential is large, $\langle \mu \rangle\sim100t$ and $\langle V_z \rangle\sim100t$, which is central to the topological protection. On the other hand, a single impurity that breaks $\mathbb{Z}_3$ symmetry, no matter how weak, is enough to lift the threefold degeneracy.

\subsection{Fock-parafermion spectral function\label{sec:FPFSpecFun}}

\begin{figure}[t]
	\begin{center}
		\includegraphics[width=1\columnwidth]{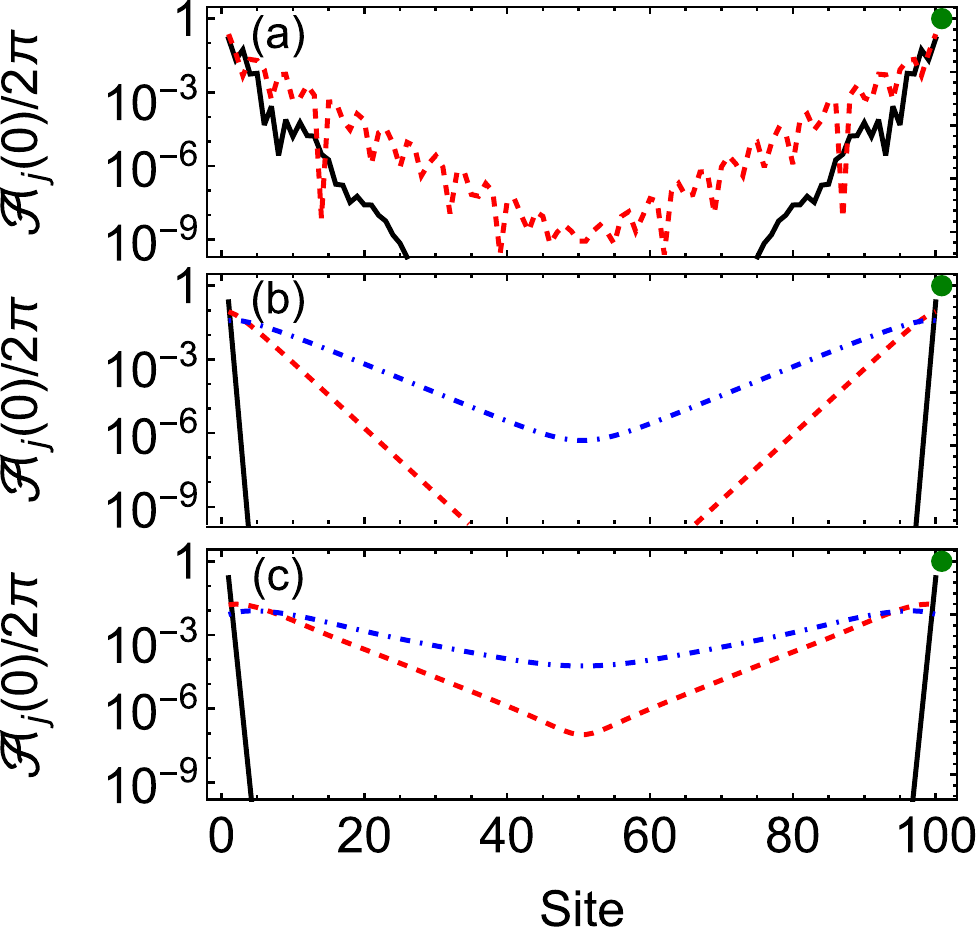}
		\caption{FPF spectral function $\mathcal{A}_j(0)$ for different interactions.  (a)  Spatial spread of the ground state over the chain for $H_{I}$ with $W_6=2.2t$ (solid black) and $W_6=3.2t$ (dashed red). (b-c) Spatial spread of the ground state of $H_{II}$ for $W_3=t$ and (b) $\mu=0.1t$ (solid black), $\mu=2.5t$ (dashed red), and $\mu=2.8t$ (dot-dashed blue) and (c) for  $V_z=0.1t$ (solid black), $V_z=2.1t$ (dashed red), and $V_z=2.2t$ (dot-dashed blue).  The green dots mark the analytical value (2/9) for perfectly localized $\mathbb{Z}_3$ edge parafermions of $H_{\rm pf}$.}
		\label{fig:SF}
	\end{center}
\end{figure}

We now turn to the spatial distribution of the parafermionic modes along the chain. To this end, we calculate the zero-energy Fock parafermion spectral function at site $j$ defined as 
\begin{align}\label{eq:FPFspec}
\mathcal{A}_j(0)\! =\! \frac{2\pi}{n_{\rm gs}}\sum_{\ket{g'}\ket{g}}\!|\!\bra{g'}d_j \bar{\omega}^{N_j}\ket{g}\!|^2 \!+\!|\!\bra{g'}\bar{\omega}^{N_j} d_j^{\dagger}\ket{g}\!|^2,   
\end{align}
where $d$ is the Fock-parafermion operator defined in Eq.~\eqref{eq:d_j} in Appendix~\ref{Appendix:Fermionization}, {$N_j=n_{\uparrow,j}+2n_{\downarrow,j}$ is the Fock-parafermion number operator} and the second sum (normalized by the ground states degeneracy $n_{\rm gs}$) runs over all ground-state {$\ket{g},\ket{g'}$}. As discussed in Appendix~\ref{supp:derivation}, the phase factor $\bar{\omega}^{N_j}$ prevents spurious asymmetries in the FPF spectral weights along the chain \cite{Mazza:ParafermionGS:2017}. Interestingly, the phase factor  $\bar{\omega}^{N_j}$ does not affect the FPF spectral function of $H_I$, Fig.~\ref{fig:SF}(a). This implies that the structure of the ground states of $H_I$ is significantly different from that of the ground states of $H_{II}$, as discussed below.

For a parafermion chain with no local interactions (Eq.~\eqref{eq:pf}), the zero-energy FPF spectral function is $\mathcal{A}_j(0)/(2 \pi) \!=\! 2/9(\delta_{j,1}\!+\!\delta_{j,L})$, which is perfectly consistent with our simulations.  In the Appendix~\ref{supp:derivation}, we show the derivation for the analytic values of $\mathcal{A}_j(0)$, a generic $\mathbb{Z}_M$ parafermion.

We emphasize that $\mathcal{A}_j(0)$ measures the local density of states related to Fock parafermions instead of the electrons, although the actual calculations involve fermionic matrix elements. A more na\"ive approach would be to calculate the purely fermionic spectral function, as it has been done  in the $\mathbb{Z}_4$ case \cite{Calzona:Z4FermionLattice:2018,Teixeira:QD:2021}. { However, the matrix elements entering the usual fermionic spectral function couple states with opposite fermionic parities and produce ill-defined results for these models. For $H_{I}$, all ground states have the same parity, such that $\bra{g'}c_\sigma\ket{g}=0$, while for $H_{II}$,  the ground states simply do not have a well-defined parity. }
{ In addition, the terms arising from bulk states do not necessarily cancel each other, which is a useful property here (see Appendix~\ref{supp:derivation}). For these reasons, $\mathcal{A}_j(0)$ as defined above is a better option to  visualize the edge parafermionic modes.}

{  The FPF spectral function for $H_{I}$ (Fig. ~\ref{fig:SF}(a)) shows exponentially localized edge states in the topological phase ($W_6 \gtrsim 2t$). These modes decay exponentially into the interior of the chain but in a nonmonotonic fashion, with an oscillation period of a few sites.  }

{ This is in sharp contrast with the strongly localized edge states of $H_{II}$ for $W_3\!=\!t$ shown in Figs.~\ref{fig:SF}(b-c). For small values of on-site potentials ($\mu$ and $V_z$, black curves), the decay occurs within a few ($\sim 10$) sites. For $\mu\!=\!V_z\!=\!0$ and $W_3\!=\!t$, $H_{II}$ maps exactly into $H_{\rm pf}$ and the parafermionic modes become free, with the FPF spectral function being zero in all sites of the chain except at the end sites, where it reaches the analytically obtained value of 2/9 (green dots in Fig. ~\ref{fig:SF}). }

{These differences between $H_{I}$ and $H_{II}$ are also encoded in the structure of the ground states in the FPF basis. For instance, while two ground states ($| g \rangle$ and $| g^{\prime} \rangle)$  of $H_{II}$ with distinct $\mathbb{Z}_3$ parity values can be coupled by any local FPF creation/destruction operator such that $\langle g | d_j + d^\dagger_j |g^\prime \rangle \neq 0$, the same is \emph{not} true for the ground states of $H_{I}$. Although the latter have well-defined $\mathbb{Z}_3$ parity values, they are not eigenstates of all local FPF number operators $n_{d,j}=d^\dagger_j d_j$.
}

 Figure~\ref{fig:SF}(b) shows the spreading of the parafermionic state as we increase the doping across the phase transition at $\mu \approx 2.9$. The plot of $\mathcal{A}_j(0)$ shows exponentially localized modes in the topological phase  ($\mu=0.5t$ (black) and $\mu = 2.5t$ (dashed red)), while the ground state becomes delocalized near the transition point ($\mu = 2.8t$ (dashed blue)). Moreover, it becomes clear that finite-size effects can be considerable in small chains ($< 100$-site long chains). 
 
 The same analysis can be done in the case of a magnetic field in the $z$ direction (Fig.~\ref{fig:SF}(b)) for $\mu=0.1 t$. Again, the spectral function shows exponentially localized edge modes for $V_z < V^{c}_z = 2.2t$, i.e., before the phase transition at $V^{c}_z = 2.2t$ (dashed blue curve), at which point the ground-state spectral function is spread all over the chain.).

\subsection{Entanglement entropy\label{sec:EntEntropy}}

\begin{figure}[t]
	\begin{center}
		\includegraphics[width=1\columnwidth]{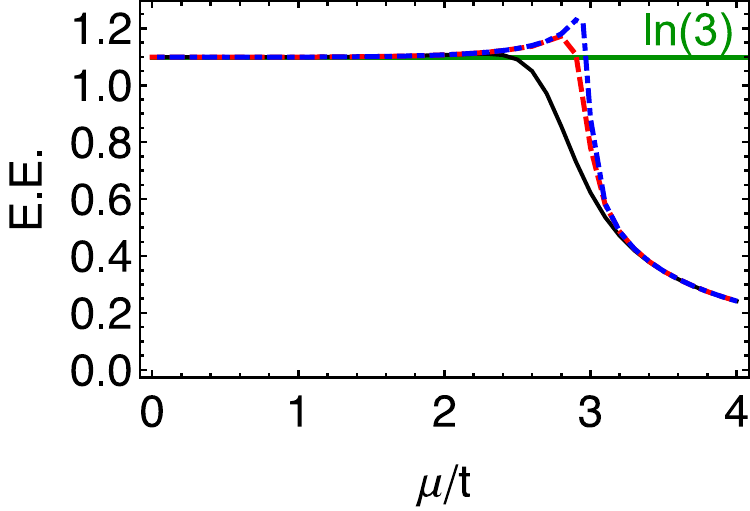}
		\caption{Entanglement entropy (EE) dependence with doping for different chain lengths of 16 (solid black), 48 (dashed red) and 100 (dotted blue). Note that far away from the phase transition all of them have the same value.} 
		\label{fig:EE}
	\end{center}
\end{figure}

Lastly, we consider the signatures of the topological phase transition in the EE \cite{Li:Phys.Rev.B:245121:2013,Zhuang:PhaseDiagramZ3Parafermion:2015}. Figure \ref{fig:EE} shows the EE. calculated at the central link of the chain as a function of the chemical potential $\mu$ for different chain sizes.

 For small values of $\mu$ such that the system is in the topological phase, EE is constant and pinned at $\ln(3)$ (main panel of  Fig.~\ref{fig:EE}). This is consistent with previous DMRG results for $\mathbb{Z}_3$ parafermion chains \cite{Zhuang:PhaseDiagramZ3Parafermion:2015}. As $\mu$ increases and the system approaches the topological phase transition, the EE increases, reaches a maximum near the phase transition, and then decreases. This behavior is accentuated for larger chains, as shown in Fig.~\ref{fig:EE}.

 \section{Concluding remarks}
 \label{sec:Conclusions}

 In conclusion, in this paper we study a family of purely one-dimensional fermionic models which map into a Kitaev-like chain of $\mathbb{Z}_3$ parafermions. Similarly to the case of Majorana zero modes, the system has a topological phase with exponentially localized $\mathbb{Z}_3$ parafermionic modes at its ends. 
 
 A key element in the proposed models is the presence of strong, Hubbard-like, repulsive interactions of strength $U_H$ on each site of the fermionic chain, effectively restricting the local Hilbert space to a $t\!-\!J$ model-like basis of zero and singly occupied (spinful) fermionic states. Within this basis, an exact the mapping of the parafermion chain to a fermionic model is obtained. Even though the mapping is exact only in the $U_H \rightarrow \infty$ limit, we show (see Appendix \ref{sec:Do}) that the parafermionic phase is present even for moderate values of the interaction in the range $U_H/t \gtrsim 10$.

Although this mapping produces nonphysical parity-breaking terms, we show that such terms can be understood as a mean-field reduction of a parity-preserving three-body interaction term. In fact, we establish that this rather exotic three-body, spin-flipping interaction term is directly responsible for the existence of a $\mathbb{Z}_3$  topological phase with  parafermionic edge modes. More importantly, we show that the ground state of the resulting Hamiltonian does have a well-defined parity and cannot be understood as a combination of symmetry-broken $\mathbb{Z}_2$ Majorana modes. 

The existence of a topological phase in the fermionic models is strongly implied by the three-fold degenerate ground states with gapless edge states and their properties such as indistinguishability by local operators, and protection against disorder and $\mathbb{Z}_3$-symmetry-preserving impurities. In this regard, we should note that similar parafermionic modes have been referred to as ``nontopological parafermions" \cite{Mazza:NonTopological:2018} or the ``poor man's parafermion" \cite{Kane:Poorman:2015} in previous studies.

 Our DMRG calculations show topological phase transitions as a function of several parameters. We show that phase transitions can be characterized by different metrics such as many-body gap closings and openings, changes in the ground-state degeneracy, and peaks in the entanglement entropy. These calculations confirm  the topological equivalency of the fermionic models and the parafermion chain and provide a way to probe the robustness of the topological phase against one-body terms such as on-site Zeeman field in the $z$ direction and changes in the chemical potential. In particular, we show that an in-plane ($xy$) magnetic field can produce phase transitions depending on the angle $\theta$ between $x$ and $y$ components. This produces a threefold anisotropy in the  energy gap and ground-state degeneracy with $\theta$ stemming from the expected $\mathbb{Z}_3$ symmetry of the original fermionic Hamiltonian. 
 
 Moreover, the FPF-SF confirms the exponential localization of the parafermionic modes deep in the topological regime {for both $H_{I}$ and $H_{II}$ models}. As the system approaches the phase transition, the FPF-SF becomes more delocalized as the edge modes located at opposite ends of the chain overlap with each other. Such finite-size effects turn out to be very relevant for small chains (less than $\sim 100$ sites).
 
Finally, while these results are an important step in understanding how $\mathbb{Z}_3$ (and more generally odd-$N$ $\mathbb{Z}_{N}$) parafermionic modes can emerge in purely fermionic systems, there are certainly several questions open. For instance, can one detect $\mathbb{Z}_{3}$ parafermionic modes using coupling to real fermions as in the $\mathbb{Z}_{4}$ case \cite{Teixeira:QD:2021}? Can these modes be used to produce topologically protected Fibonacci anyons \cite{Stoudenmire:AssemblingFibonacci:2015}? Do these modes have different braiding statistics from pure (nonfermionic) $\mathbb{Z}_{3}$ parafermions, similar to $\mathbb{Z}_4$ parafermions \cite{Alicea:FermionezParafermionsSEMajorana:2018}? These are all relevant questions which should be explored in future works on the topic.

\acknowledgments

We thank Thomas Schmidt for enlightening discussions. We acknowledge financial support from Brazilian agencies FAPESP (Grants no.  2019/11550-8 and 2021/07602-2 ), Capes, and CNPq (Graduate scholarship program 141556/2018-8, and Research Grants 308351/2017-7, 423137/2018-2, and 309789/2020-6).

\appendix
%
 
 \section{Fermionization\label{Appendix:Fermionization}}
  In order to find a representation of the {parafermionic} Hamiltonian, {$H_{pf}=-J\sum_{j=1}^{L-1} \psi_j\chi^\dagger_{j+1} +H.c.$,} (Eq.~\eqref{eq:pf}), and its {dangling} parafermions in terms of fermionic operators, it is useful to consider FFPF operators \cite{Cobanera:FockParafermion:2014,Calzona:Z4FermionLattice:2018}. These operators act in the space of states with a well-defined Fock-parafermion number. Each parafermion can be described in terms of creation ($d^\dagger$) or annihilation ($d$) operators, which, respectively, increase and decrease the FPF number, as:
  \begin{align}\label{eq:FPF}
  \psi_j =  \ d_j \omega^{N_j} + d^{\dagger 2}_j && \chi_{j}=\  d_{j} + d^{\dagger 2}_{j}, \end{align}
  where $N_j = d_j^\dagger d_j  + d_j^{\dagger 2} d^2_j$ is the number of FPFs and can be either 0, 1 or 2. Because of Eq.~\eqref{eq:FPF}, FPF operators must satisfy relations similar to parafermions \cite{Cobanera:FockParafermion:2014};
  \begin{align}
  \label{eq:FPFconstraint}
  d_j d_l = \omega \ d_l d_j, & \hspace{0.5cm} d_j d_l^\dagger = \omega \ d_l^\dagger d_j,  \hspace{0.5cm}\text{for } l<j \nonumber\\
  &d^3=d^{\dagger 3}=0\\
  &\hspace{-2cm}d^{\dagger m}_j d^m_j + d^{3-m}_j d^{\dagger 3-m}_j =1  \text{\hspace{0.5cm}for } m= 1,2\nonumber
  \end{align}

  In order to represent operator $d$ in a fermionic representation, we choose a mapping between FPF number and fermionic number basis such that each state in the $t \!-\! J$ fermionic basis \cite{Batista:GeneralizedJW:2001} ($|\text{E}\rangle,c^\dagger_{\uparrow} |\text{E}\rangle,$ and $c^\dagger_{\downarrow} |\text{E}\rangle$, with $|\text{E}\rangle$ a vacuum state) corresponds to one state in the FPF number basis ($|0\rangle,|1\rangle,|2\rangle$). This mapping can be summarized as 
  \begin{gather}
  |2\rangle \xrightarrow{\mathmakebox[0.03\columnwidth]{\text{\large $d$}}} |1\rangle \xrightarrow{\mathmakebox[0.03\columnwidth]{\text{\large $d$}}} |0\rangle \xrightarrow{\mathmakebox[0.03\columnwidth]{\text{\large $d$}}} \emptyset \nonumber \\\label{eq:maps} \\
  c^{\dagger}_{\downarrow}|\text{E}\rangle \xrightarrow{\mathmakebox[0.03\columnwidth]{\text{\large $d$}}} c^{\dagger}_{\uparrow}|\text{E}\rangle \xrightarrow{\mathmakebox[0.03\columnwidth]{\text{\large $d$}}} |\text{E}\rangle \xrightarrow{\mathmakebox[0.03\columnwidth]{\text{\large $d$}}} \emptyset. \nonumber
  \end{gather}
  
  With this in mind, it is straightforward to a find a representation of FPF operators, $d\!=\!c_{\uparrow} \!+\! c^\dagger_{\uparrow}c_{\downarrow}$. It is also easy to see that Eq~\eqref{eq:maps} satisfies all FPF operator relations. However, since the FPF operators operate in real space (sites in a chain), we need to consider Jordan-Wigner (JW) string factors for both FPF operators and fermionic operators \cite{Cobanera:FockParafermion:2014,Calzona:Z4FermionLattice:2018}:
  \begin{align}\label{eq:d_j}
  d_j = \omega^{\sum_{p<j}N_p}\left[(-1)^{\sum_{p<j}n_p}c_{\uparrow,j} + c^\dagger_{\uparrow,j}c_{\downarrow,j}\right],\\ 
  \nonumber \\
\label{eq:d_j2}  d_j^2 =  \omega^{\sum_{p<j}2N_p}\left[(-1)^{\sum_{p<j}n_p}c_{\downarrow,j}\right],
  \end{align}
  where $n_p$ is the occupation number ($0$ or $1$) of site $p$ and $N_p \!=\! n_{\uparrow,p} \!+\!2 n_{\downarrow,p}$. 
  
  An important consequence of these strings is that FPF operators have two distinct long-range behaviors, namely (i) a JW-like string 
  which depends on the FPF number $N_p$
  applied uniformly to terms and (ii) a JW string which depends on the fermionic occupation $n_p$ applied only on single fermion operators. This is more distinct than in the case of $\mathbb{Z}_4$ parafermions where all terms have the same parity and string factor \footnote{The problem arises in the $\mathbb{Z}_3$ case mainly because there is no single fermionic operator connecting the up and down states. By contrast, in the case of $\mathbb{Z}_4$ parafermions, two sequential FPF numbers are connected by a single fermionic operator, eliminating the problem.}. This means that, apart from the expected Jordan-Wigner string, the Hamiltonian is  local in the FPF space (although nonlocal in the fermionic basis), allowing one to derive local quantities that identify the edge states. 
 
 {The parafermion operators $\chi_j$ and $\psi_j$ can be easily written in terms of the usual fermionic operators as
  \begin{align}
  \label{eq:dangling}
  \chi_j=&\  \omega^{\sum\limits_{\hspace{0.1cm}p<j}\hspace{-0.1cm}N_p} \left[(-1)^{\sum\limits_{\hspace{0.1cm}p<j}\hspace{-0.1cm}n_p}[c_{\uparrow,j} +  c^\dagger_{\downarrow,j}] + c^\dagger_{\uparrow,j} c_{\downarrow,j}\right],\\
  \psi_j = & \ \omega^{\sum\limits_{\hspace{0.1cm}p<j}\hspace{-0.1cm}N_p} \left[(-1)^{\sum\limits_{\hspace{0.1cm}p<j}\hspace{-0.1cm}n_p}[\omega c_{\uparrow,j}+  c^\dagger_{\downarrow,j}] + \omega^2 c^\dagger_{\uparrow,j} c_{\downarrow,j}\right] \nonumber \; .
  \end{align}
}

{  In particular,  we have the dangling parafermion modes written as:
 \begin{align}
  \label{eq:dangling1L}
    \chi_1 =& c_{\uparrow,1} \!+\!  c^\dagger_{\downarrow,1} \!+\! c^\dagger_{\uparrow,1} c_{\downarrow,1}, \\
     \psi_L =& \omega^{\sum\limits_{\hspace{0.1cm}p<L}\hspace{-0.1cm}N_p}  \left[(-1)^{\sum\limits_{\hspace{0.1cm}p<L}\hspace{-0.1cm}n_p} [\omega c_{\uparrow,L}\!+\!  c^\dagger_{\downarrow,L}] \!+\! \omega^2 c^\dagger_{\uparrow,L} c_{\downarrow,L}\right] \nonumber \;. 
   \end{align}
}
   
     Notice that $\psi_L$ contains information of the fermionic occupation in the central chain sites. This means that the edge modes are affected by the bulk states via Jordan-Wigner strings. While this also occurs in the $\mathbb{Z}_4$ case \cite{Calzona:Z4FermionLattice:2018}, the difference here is that the strings are not applied uniformly in every term, due to absence of a well-defined parity of the operators. 
  
  Using the above relations~\eqref{eq:dangling}, we can express the Hamiltonian $H_{pf}$ in terms of fermionic operators:
  \begin{align}
      H&_{pf}=-J\sum_{j=1}^{L-1} \psi_j\chi^\dagger_{j+1} +H.c.\nonumber\\
      =&-J\sum_{j=1}^{L-1} \left[(-1)^{\sum\limits_{\hspace{0.1cm}p<j}\hspace{-0.1cm}n_p}[c_{\uparrow,j}\!-\! i c^\dagger_{\downarrow,j}] \!+\! i c^\dagger_{\uparrow,j} c_{\downarrow,j}\right]\! \\
      \times&\left[(-1)^{\sum\limits_{\hspace{0.1cm}p<j+1}\hspace{-0.1cm}n_p}[c^\dagger_{\uparrow,j+1} \!+  \!i c_{\downarrow,j+1}] \!-\! i c^\dagger_{\downarrow,j+1} c_{\uparrow,j+1}\right]\!+\!H.c.,\nonumber
  \end{align}
which is equal to $H_{II}$ when $t\!=\!\Delta\!=\!W_3\!=\!W_4\!=\!J$.

 \section{Mean-field derivation\label{Appendix:MF}}
In this appendix, we use mean-field arguments to obtain an expression similar  to Eq.~\eqref{eq:H(3)} starting from Eq.~\eqref{eq:H(6)}. 

We start with the spin-up terms in Eq.~\eqref{eq:H(6)}:
\begin{align}
\label{eq:spinupterms}
c^\dagger_{\uparrow,i-1}&c_{\downarrow,i-1}c^\dagger_{\uparrow,i}c_{\downarrow,i}c^\dagger_{\uparrow,i+1}c_{\downarrow,i+1}\approx\\
&\langle c^\dagger_{\uparrow,i-1}c_{\downarrow,i-1}c^\dagger_{\uparrow,i} \rangle  c_{\downarrow,i}c^\dagger_{\uparrow,i+1}c_{\downarrow,i+1}+\nonumber\\
&\langle c^\dagger_{\uparrow,i-1}c_{\downarrow,i-1}c_{\downarrow,i}\rangle c^\dagger_{\uparrow,i}c^\dagger_{\uparrow,i+1}c_{\downarrow,i+1}+\nonumber\\
&c^\dagger_{\uparrow,i-1}c_{\downarrow,i-1}c^\dagger_{\uparrow,i}\langle c_{\downarrow,i}c^\dagger_{\uparrow,i+1}c_{\downarrow,i+1}\rangle+\nonumber\\
&c^\dagger_{\uparrow,i-1}c_{\downarrow,i-1}c_{\downarrow,i}\langle c^\dagger_{\uparrow,i}c^\dagger_{\uparrow,i+1}c_{\downarrow,i+1}\rangle\nonumber
\end{align}
where we use the commutation relation $[c^\dagger_{\uparrow,i},c_{\downarrow,j}]\!=\!\delta_{i,j}c^\dagger_{\uparrow,i}c_{\downarrow,i}$ arising from the $t\!-\!J$ model requirement of exclusion of double occupancy states. 

Assuming a spatially isotropic and $SU(2)$-symmetric spin, the expectation values in Eq.~\eqref{eq:spinupterms} should be proportional to $(-1)^{\sum_{p<i} n_p}$. Indeed this can be seen by calculating the expectation value $\langle c^\dagger_{\uparrow,i-1}c_{\downarrow,i-1}c^\dagger_{\uparrow,i} \rangle$ for the ground state of $H_{II}$ in the case of $t=\Delta=W_3=W_4$ (when it is mapped exactly to $H_{pf}$). In order to compute the trifermion expectation value, we go to the Fock-parafermion basis:
\begin{equation}\label{eq:expectation}
    \langle c^\dagger_{\uparrow,j-1}c_{\downarrow,j-1}c^\dagger_{\uparrow,j} \rangle\!=\!\langle (-1)^{\sum\limits_{\hspace{0.1cm}p<j}\hspace{-0.1cm}n_p} \!d^\dagger_{j-1}d^{2}_{j-1}\omega^{N_j}d_j d^{\dagger 2}_j \rangle,
\end{equation}
and since we are at zero temperature, the expectation value will be the average over the ground-states. We can check if this expectation value can be different from zero by considering the $\mathbb{Z}_3$ parafermion ground state of $H_{pf}$ with $L$ sites and total Fock-parafermion number $i$ given by~\cite{Mazza:ParafermionGS:2017}
\begin{equation}
\label{eq:3gs}
|g_i^L\rangle  = \frac{1} {\sqrt{3^{L-1}}}\sum_{\substack{\{N_j\} \text{ such that} \\ \sum_j N_j =i \ mod \ 3}} \bigotimes_{j=1}^L|N_j\rangle,
\end{equation}
which, in turn, can be written in terms of a chain with $L\!-\!2$ sites and two sites at one of the ends or two sites at position $s$ and $s+1$ together with two chains with $s\!-\!2$ and $L\!-\!s-1$ sites:
\begin{gather}
|g_i\rangle  = \frac{1}{\sqrt{3}}\sum_{k=0}^{2}\sum_{l=0}^{2}\ket{f_{i-k}}\otimes\ket{f_{k-l}}\otimes\ket{g_l^{L-2}}\nonumber\\
\label{eq:3gsSplit}
|g_i\rangle  = \frac{1}{\sqrt{3}}\sum_{k=0}^{2}\sum_{l=0}^{2}\ket{g_l^{L-2}}\otimes\ket{f_{i-k}}\otimes\ket{f_{k-l}}\\
|g_i\rangle  = \frac{1}{3}\sum_{a=0}^{2}\sum_{k=0}^{2}\sum_{l=0}^{2}\ket{g_a^{s-2}}\otimes\ket{f_{i-k}}\otimes\ket{f_{k-l}}\otimes\ket{g_{-a+l}^{L-s}},\nonumber
\end{gather} 
where $\ket{f_i}$ is the state of a single site with FPF number $f_i$, such that  $\bra{f_j}|d^k\ket{f_i} \!=\! \delta_{j,i-k}$ for $k\!\leq\! i$ and zero otherwise. When we apply $d^\dagger_{j-1}d^{2}_{j-1}\omega^{N_j}d_j d^{\dagger 2}_j$ on $\ket{g_i}$ the only nonzero terms are sums with $i\!-\!k\!=\!2$ and $k\!-\!l\!=\!0$. Because of the structure of the ground state we can compute the expectation value of Eq.~\eqref{eq:expectation}:
\begin{align}\label{eq:expectationCalc}
    \bra{g_i} &(-1)^{\sum\limits_{\hspace{0.1cm}p<j}\hspace{-0.1cm}n_p} \!d^\dagger_{j-1}d^{2}_{j-1}\omega^{N_j}d_j d^{\dagger 2}_j\ket{g_i}\\
    =&\frac{1}{9}\sum_{a,k,l}\bra{g_a^{j-2}}(-1)^{\sum\limits_{\hspace{0.1cm}p<j-1}\hspace{-0.1cm}n_p}\ket{g_a^{j-2}}\bra{f_{i-k}}d^\dagger d^{2}(-1)^n\ket{f_2}\nonumber\\
    &\hspace{5cm}\times\bra{f_{k-l}} \omega^{N}d d^{\dagger 2}\ket{0} \nonumber\\
    &=-\frac{\omega}{9}\sum_{a}\bra{g_a^{j-2}}(-1)^{\sum\limits_{\hspace{0.1cm}p<j-1}\hspace{-0.1cm}n_p}\ket{g_a^{j-2}}=\frac{\omega (-1)^{j-1}}{3^{j}}\nonumber
\end{align}
where we used Eq.~\eqref{eq:3gs} to calculate the sum of the string factors, $\sum_{a}\bra{g_a^{j-2}}(-1)^{\sum\limits_{\hspace{0.1cm}p<j-1}\hspace{-0.1cm}n_p}\ket{g_a^{j-2}}=(-1/3)^{j-2}$. This means that this correlation decays away from the first site $j\!=\!1$. This does not imply that there is no parafermion in $H_I$ as we consider only one way of pairing the operators. To recover the expression of $H_{3}$, we need to substitute the average value of the string by its operator, $\langle (-1)^{\sum\limits_{\hspace{0.1cm}p<j}\hspace{-0.1cm}n_p}\rangle\to (-1)^{\sum\limits_{\hspace{0.1cm}p<j}\hspace{-0.1cm}n_p}$, and we obtain

\begin{align}
H^{(6)}_{MF} \!=\! & -W_{MF} \sum_ {j}^{L-1} (-1)^{\sum\limits_{\hspace{0.1cm}p<j}\hspace{-0.1cm}n_p} \Big [( c_{\uparrow,j} +  c^\dagger_{\downarrow j}) c^\dagger_{\downarrow,j+1}c_{\uparrow,j+1}\nonumber\\ 
&\hspace{1cm}+ \ c^\dagger_{\uparrow,j} c_{\downarrow,j} ( c^\dagger_{\uparrow,j+1} +  c_{\downarrow,j+1})\Big]+ H.c.
\end{align}
which is the similar to Equation \eqref{eq:H(3)} for an infinite chain, i.e., without edges.

\section{Allowing double occupancy}\label{sec:Do}

In the previous sections, we considered a local fermionic basis excluding the double occupancy state, $c_{\downarrow}^\dagger c_{\uparrow}^\dagger\ket{E}$. A consistency check for this approach would be to include this state in the fermionic basis along with a Hubbard interaction in each site, $U_{H}n_\uparrow n_\downarrow$, and then take the limit  $U_{H}\to\infty$. In this Appendix, we perform  this consistency check and show that indeed we recover the main text's results.

\begin{figure}[t]
	\begin{center}
		\includegraphics[width=1\columnwidth]{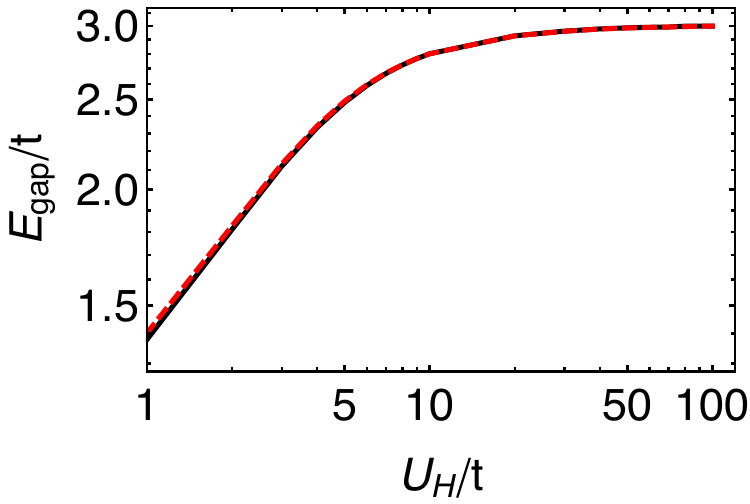}
		\caption{Gap dependency with Hubbard interaction in a double occupancy basis. The gaps of the 16-site chain (red dashed) and 100-site chain  (solid black) have minor differences only in the low interaction regime, $U_H\approx t$.} 
		\label{fig:DoubleO}
	\end{center}
\end{figure}

Figure \ref{fig:DoubleO} illustrates the persistence of the parafermionic phase already for relatively small values of $U_{H}$ relative to the hopping $t$ (say, $U_{H}/t \sim 1-5$). For $U_{H}\gg t$, the gap becomes completely independent of the chain size, and we recover the expected $E_{\rm gap}/t\!=\!3$. 
This shows that a large Hubbard (on-site) interaction is not necessary for the formation of $\mathbb{Z}_3$ parafermions and that finite-size effects are not relevant in this regime.

\section{FPF spectral function derivation}
\label{supp:derivation}

In this section, we show that the FPF spectral function for a $\mathbb{Z}_M$ parafermion chain with two dangling parafermions described by Eq.~\eqref{eq:pf}  is given by  $\mathcal{A}_j \!=\! 2\pi\frac{2}{M^2}\left( \delta_{1,j} \!+\!\delta_{L,j}\right)$. In this appendix only we consider $\omega=e^{2 i \pi/M}$. In this section we use a  generalization of the ground state shown in Appendix~\ref{Appendix:MF} for $M\geq 2$.

We start with the FPF spectral function, defined by Eq.~\eqref{eq:FPFspec}. We can write it as
\begin{align}
\mathcal{A}_j(E') =& \frac{2\pi}{N_{gs}}\sum_{\ket{\varphi}\ket{g}} \delta(E'+E_\varphi-E_0)|\bra{\varphi}d_j \bar{\omega}^{N_j}\ket{g}|^2 \nonumber\\&+\delta(E'- E_\varphi+E_0)|\bra{\varphi}\bar{\omega}^{N_j} d_j^{\dagger}\ket{g}|^2,   
\end{align}
where we sum over all ground states $\ket{g}$ and divide by its degeneracy $n_{gs}$. The state $\ket{\varphi}$ is an eigenstate of the Hamiltonian with energy $E_{\varphi}$. As we show later, it is important to consider $d_j \bar{\omega}^{N_j}$ \footnote{Another option is to consider $d_j \bar{\omega}^{N_j/2}$ which will also have a symmetric zero-energy spectral function. The biggest difference is a scaling factor and should not affect how the edge states are localized.} instead of just $d_j$ due to symmetry of the spectral function. While the former has a symmetric zero-energy spectral function along the chain, the latter will have the zero-energy spectral function on the first site, $j\!=\!1$, $(M-1)^2$ times larger than on the last site, $j\!=\!L$.


For a generic chain with $L$ sites, the ground state of $H_{pf}$ for $\mathbb{Z}_M$ parafermions with total FPF number $i$~\cite{Mazza:ParafermionGS:2017} is given by
\begin{equation}
\label{eq:gs}
|g_i^L\rangle  = \frac{1}{\sqrt{M^{L-1}}}\sum_{\substack{\{N_j\} \text{ such that} \\ \sum_j N_j =i \ mod \ M}} \bigotimes_{j=1}^L|N_j\rangle,
\end{equation}
which, in turn, can be written in terms of a chain with $L\!-\!1$ sites and a single site at one of the ends or a single site at position $s$ and two chains with $s\!-\!1$ and $L\!-\!s$ sites:
\begin{gather}
\label{eq:gsLeft}
|g_i\rangle  = \frac{1}{\sqrt{M}}\sum_{k=0}^{M-1}\ket{f_{i-k}}\otimes\ket{g_k^{L-1}}\\
\label{eq:gsRight}
|g_i\rangle  = \frac{1}{\sqrt{M}}\sum_{k=0}^{M-1}\ket{g_k^{L-1}}\otimes\ket{f_{i-k}}\\
\label{eq:gsMid}
|g_i\rangle  = \frac{1}{M}\sum_{a=0}^{M-1}\sum_{k=0}^{M-1}\ket{g_a^{s-1}}\otimes\ket{f_{i-k}}\otimes\ket{g_{-a+k}^{L-s}},
\end{gather} 

It is straightforward to see that at positions $j\!=\! 1,L$ we have a sum over $M-1$ different powers of $\omega$, and its absolute value is always $1$. This leads to $ |\bra{g_{i-1}^L}d_j\bar{\omega}^N_j\ket{g_i^L}|\!=\!1/M^2$. In the bulk, Eq.~\eqref{eq:gsMid}, we need to consider the FPF commutation relations, Eq.~\eqref{eq:FPFconstraint}.

Applying $d_j$ at $|g_i\rangle$, with $j$ in the bulk, does not only decrease the FPF number in one but also adds a phase that depends on the FPF number that precedes it,:
\begin{align}
d_j \bar{\omega}^N|g_i\rangle & = \frac{1}{M}\sum_{a=0}^{M-1}\sum_{k=0}^{M-1} \omega^{a-i+k+1}\ket{g_a^{j-1}}\nonumber\\&\hspace{2cm}\otimes d\ket{f_{i-k}}\otimes\ket{g_{-a+k}^{L-j}},
\end{align}
such that $\bra{g_{i-1}} d_j \bar{\omega}^{N_j} \ket{g_i} \!\propto \!\sum_{a=0}^{M-1} e^{2 \pi i a /M} \!=\!0$. The same procedure can be done for $\bar{\omega}^N_j d^\dagger_j$ and yields the same result. Therefore, the spectral function, at zero energy, of a $\mathbb{Z}_M$ parafermion chain is given by
\begin{align}
\mathcal{A} = \frac{2\pi}{N_{gs}}\sum_{\ket{\varphi}\ket{g}}&|\bra{\varphi}d_j \bar{\omega}^{N_j}\ket{g}|^2 +|\bra{\varphi}\bar{\omega}^{N_j} d_j^{\dagger}\ket{g}|^2,\nonumber\\
& \hspace{1cm} =2\pi\frac{2}{M^2}\left( \delta_{1,j} +\delta_{L,j}\right)
\end{align}

We also note that using only $d_j$ instead of $d_j \bar{\omega}^{N_j}$ leads to an asymmetry between sites 1 and $L$ which does not make sense in terms of how the parafermions are localized. This fact is unrelated with the fermionic basis (and its long-range interaction) and happens in ``pure'' parafermion models \cite{Mazza:ParafermionGS:2017}. In addition, the phase factor $\bar{\omega}^N_j$ is not unique. A factor such as $\omega^{N_j/2}$ would also work, albeit it introduces an additional scale factor related with the parafermion chain length. Other powers of $\omega^N_j$ might also work but they are not universal, i.e., they would depend on the value of $M$ of the $\mathbb{Z}_M$ parafermion.


\begin{thebibliography}{54}%
\makeatletter
\providecommand \@ifxundefined [1]{%
 \@ifx{#1\undefined}
}%
\providecommand \@ifnum [1]{%
 \ifnum #1\expandafter \@firstoftwo
 \else \expandafter \@secondoftwo
 \fi
}%
\providecommand \@ifx [1]{%
 \ifx #1\expandafter \@firstoftwo
 \else \expandafter \@secondoftwo
 \fi
}%
\providecommand \natexlab [1]{#1}%
\providecommand \enquote  [1]{``#1''}%
\providecommand \bibnamefont  [1]{#1}%
\providecommand \bibfnamefont [1]{#1}%
\providecommand \citenamefont [1]{#1}%
\providecommand \href@noop [0]{\@secondoftwo}%
\providecommand \href [0]{\begingroup \@sanitize@url \@href}%
\providecommand \@href[1]{\@@startlink{#1}\@@href}%
\providecommand \@@href[1]{\endgroup#1\@@endlink}%
\providecommand \@sanitize@url [0]{\catcode `\\12\catcode `\$12\catcode
  `\&12\catcode `\#12\catcode `\^12\catcode `\_12\catcode `\%12\relax}%
\providecommand \@@startlink[1]{}%
\providecommand \@@endlink[0]{}%
\providecommand \url  [0]{\begingroup\@sanitize@url \@url }%
\providecommand \@url [1]{\endgroup\@href {#1}{\urlprefix }}%
\providecommand \urlprefix  [0]{URL }%
\providecommand \Eprint [0]{\href }%
\providecommand \doibase [0]{https://doi.org/}%
\providecommand \selectlanguage [0]{\@gobble}%
\providecommand \bibinfo  [0]{\@secondoftwo}%
\providecommand \bibfield  [0]{\@secondoftwo}%
\providecommand \translation [1]{[#1]}%
\providecommand \BibitemOpen [0]{}%
\providecommand \bibitemStop [0]{}%
\providecommand \bibitemNoStop [0]{.\EOS\space}%
\providecommand \EOS [0]{\spacefactor3000\relax}%
\providecommand \BibitemShut  [1]{\csname bibitem#1\endcsname}%
\let\auto@bib@innerbib\@empty
\bibitem [{\citenamefont {Nayak}\ \emph {et~al.}(2008)\citenamefont {Nayak},
  \citenamefont {Simon}, \citenamefont {Stern}, \citenamefont {Freedman},\ and\
  \citenamefont {Das~Sarma}}]{Nayak:NonAbelianAnyonTQC:2008}%
  \BibitemOpen
  \bibfield  {author} {\bibinfo {author} {\bibfnamefont {C.}~\bibnamefont
  {Nayak}}, \bibinfo {author} {\bibfnamefont {S.~H.}\ \bibnamefont {Simon}},
  \bibinfo {author} {\bibfnamefont {A.}~\bibnamefont {Stern}}, \bibinfo
  {author} {\bibfnamefont {M.}~\bibnamefont {Freedman}},\ and\ \bibinfo
  {author} {\bibfnamefont {S.}~\bibnamefont {Das~Sarma}},\ }\bibfield  {title}
  {\bibinfo {title} {Non-abelian anyons and topological quantum computation},\
  }\href {https://doi.org/10.1103/RevModPhys.80.1083} {\bibfield  {journal}
  {\bibinfo  {journal} {Rev. Mod. Phys.}\ }\textbf {\bibinfo {volume} {80}},\
  \bibinfo {pages} {1083} (\bibinfo {year} {2008})}\BibitemShut {NoStop}%
\bibitem [{\citenamefont {Kitaev}(2001)}]{Kitaev:P.U:2001}%
  \BibitemOpen
  \bibfield  {author} {\bibinfo {author} {\bibfnamefont {A.~Y.}\ \bibnamefont
  {Kitaev}},\ }\bibfield  {title} {\bibinfo {title} {Unpaired majorana fermions
  in quantum wires},\ }\href {http://stacks.iop.org/1063-7869/44/i=10S/a=S29}
  {\bibfield  {journal} {\bibinfo  {journal} {Physics-Uspekhi}\ }\textbf
  {\bibinfo {volume} {44}},\ \bibinfo {pages} {131} (\bibinfo {year}
  {2001})}\BibitemShut {NoStop}%
\bibitem [{\citenamefont {Aasen}\ \emph {et~al.}(2016)\citenamefont {Aasen},
  \citenamefont {Hell}, \citenamefont {Mishmash}, \citenamefont {Higginbotham},
  \citenamefont {Danon}, \citenamefont {Leijnse}, \citenamefont {Jespersen},
  \citenamefont {Folk}, \citenamefont {Marcus}, \citenamefont {Flensberg},\
  and\ \citenamefont {Alicea}}]{Alicea:MilestoneMajoranaQC:2016}%
  \BibitemOpen
  \bibfield  {author} {\bibinfo {author} {\bibfnamefont {D.}~\bibnamefont
  {Aasen}}, \bibinfo {author} {\bibfnamefont {M.}~\bibnamefont {Hell}},
  \bibinfo {author} {\bibfnamefont {R.~V.}\ \bibnamefont {Mishmash}}, \bibinfo
  {author} {\bibfnamefont {A.}~\bibnamefont {Higginbotham}}, \bibinfo {author}
  {\bibfnamefont {J.}~\bibnamefont {Danon}}, \bibinfo {author} {\bibfnamefont
  {M.}~\bibnamefont {Leijnse}}, \bibinfo {author} {\bibfnamefont {T.~S.}\
  \bibnamefont {Jespersen}}, \bibinfo {author} {\bibfnamefont {J.~A.}\
  \bibnamefont {Folk}}, \bibinfo {author} {\bibfnamefont {C.~M.}\ \bibnamefont
  {Marcus}}, \bibinfo {author} {\bibfnamefont {K.}~\bibnamefont {Flensberg}},\
  and\ \bibinfo {author} {\bibfnamefont {J.}~\bibnamefont {Alicea}},\
  }\bibfield  {title} {\bibinfo {title} {Milestones toward majorana-based
  quantum computing},\ }\href {https://doi.org/10.1103/PhysRevX.6.031016}
  {\bibfield  {journal} {\bibinfo  {journal} {Phys. Rev. X}\ }\textbf {\bibinfo
  {volume} {6}},\ \bibinfo {pages} {031016} (\bibinfo {year}
  {2016})}\BibitemShut {NoStop}%
\bibitem [{\citenamefont {Lutchyn}\ \emph {et~al.}(2018)\citenamefont
  {Lutchyn}, \citenamefont {Bakkers}, \citenamefont {Kouwenhoven},
  \citenamefont {Krogstrup}, \citenamefont {Marcus},\ and\ \citenamefont
  {Oreg}}]{Lutchyn:MajoranaReview:2018}%
  \BibitemOpen
  \bibfield  {author} {\bibinfo {author} {\bibfnamefont {R.~M.}\ \bibnamefont
  {Lutchyn}}, \bibinfo {author} {\bibfnamefont {E.~P. A.~M.}\ \bibnamefont
  {Bakkers}}, \bibinfo {author} {\bibfnamefont {L.~P.}\ \bibnamefont
  {Kouwenhoven}}, \bibinfo {author} {\bibfnamefont {P.}~\bibnamefont
  {Krogstrup}}, \bibinfo {author} {\bibfnamefont {C.~M.}\ \bibnamefont
  {Marcus}},\ and\ \bibinfo {author} {\bibfnamefont {Y.}~\bibnamefont {Oreg}},\
  }\bibfield  {title} {\bibinfo {title} {Majorana zero modes in
  superconductor--semiconductor heterostructures},\ }\href
  {https://doi.org/10.1038/s41578-018-0003-1} {\bibfield  {journal} {\bibinfo
  {journal} {Nature Reviews Materials}\ }\textbf {\bibinfo {volume} {3}},\
  \bibinfo {pages} {52} (\bibinfo {year} {2018})}\BibitemShut {NoStop}%
\bibitem [{\citenamefont {Flensberg}\ \emph {et~al.}(2021)\citenamefont
  {Flensberg}, \citenamefont {von Oppen},\ and\ \citenamefont
  {Stern}}]{Flensberg:ReviewMajorana:2021}%
  \BibitemOpen
  \bibfield  {author} {\bibinfo {author} {\bibfnamefont {K.}~\bibnamefont
  {Flensberg}}, \bibinfo {author} {\bibfnamefont {F.}~\bibnamefont {von
  Oppen}},\ and\ \bibinfo {author} {\bibfnamefont {A.}~\bibnamefont {Stern}},\
  }\bibfield  {title} {\bibinfo {title} {Engineered platforms for topological
  superconductivity and majorana zero modes},\ }\href
  {https://doi.org/10.1038/s41578-021-00336-6} {\bibfield  {journal} {\bibinfo
  {journal} {Nature Reviews Materials}\ }\textbf {\bibinfo {volume} {6}},\
  \bibinfo {pages} {944} (\bibinfo {year} {2021})}\BibitemShut {NoStop}%
\bibitem [{\citenamefont {Stern}\ and\ \citenamefont
  {Lindner}(2013)}]{Stern:TQC:2013}%
  \BibitemOpen
  \bibfield  {author} {\bibinfo {author} {\bibfnamefont {A.}~\bibnamefont
  {Stern}}\ and\ \bibinfo {author} {\bibfnamefont {N.~H.}\ \bibnamefont
  {Lindner}},\ }\bibfield  {title} {\bibinfo {title} {Topological quantum
  computation{\textemdash}from basic concepts to first experiments},\ }\href
  {https://doi.org/10.1126/science.1231473} {\bibfield  {journal} {\bibinfo
  {journal} {Science}\ }\textbf {\bibinfo {volume} {339}},\ \bibinfo {pages}
  {1179} (\bibinfo {year} {2013})}\BibitemShut {NoStop}%
\bibitem [{\citenamefont {Sarma}\ \emph {et~al.}(2015)\citenamefont {Sarma},
  \citenamefont {Freedman},\ and\ \citenamefont
  {Nayak}}]{Nayak:MajoranaQuantumComputation:2015}%
  \BibitemOpen
  \bibfield  {author} {\bibinfo {author} {\bibfnamefont {S.~D.}\ \bibnamefont
  {Sarma}}, \bibinfo {author} {\bibfnamefont {M.}~\bibnamefont {Freedman}},\
  and\ \bibinfo {author} {\bibfnamefont {C.}~\bibnamefont {Nayak}},\ }\bibfield
   {title} {\bibinfo {title} {Majorana zero modes and topological quantum
  computation},\ }\href {https://doi.org/10.1038/npjqi.2015.1} {\bibfield
  {journal} {\bibinfo  {journal} {Npj Quantum Information}\ }\textbf {\bibinfo
  {volume} {1}},\ \bibinfo {pages} {15001 EP } (\bibinfo {year} {2015})},\
  \bibinfo {note} {review Article}\BibitemShut {NoStop}%
\bibitem [{\citenamefont {Stoudenmire}\ \emph {et~al.}(2015)\citenamefont
  {Stoudenmire}, \citenamefont {Clarke}, \citenamefont {Mong},\ and\
  \citenamefont {Alicea}}]{Stoudenmire:AssemblingFibonacci:2015}%
  \BibitemOpen
  \bibfield  {author} {\bibinfo {author} {\bibfnamefont {E.~M.}\ \bibnamefont
  {Stoudenmire}}, \bibinfo {author} {\bibfnamefont {D.~J.}\ \bibnamefont
  {Clarke}}, \bibinfo {author} {\bibfnamefont {R.~S.~K.}\ \bibnamefont
  {Mong}},\ and\ \bibinfo {author} {\bibfnamefont {J.}~\bibnamefont {Alicea}},\
  }\bibfield  {title} {\bibinfo {title} {Assembling fibonacci anyons from a
  ${\mathbb{z}}_{3}$ parafermion lattice model},\ }\href
  {https://doi.org/10.1103/PhysRevB.91.235112} {\bibfield  {journal} {\bibinfo
  {journal} {Phys. Rev. B}\ }\textbf {\bibinfo {volume} {91}},\ \bibinfo
  {pages} {235112} (\bibinfo {year} {2015})}\BibitemShut {NoStop}%
\bibitem [{\citenamefont {Hutter}\ and\ \citenamefont
  {Loss}(2016)}]{Loss:QCwithParafermion:2016}%
  \BibitemOpen
  \bibfield  {author} {\bibinfo {author} {\bibfnamefont {A.}~\bibnamefont
  {Hutter}}\ and\ \bibinfo {author} {\bibfnamefont {D.}~\bibnamefont {Loss}},\
  }\bibfield  {title} {\bibinfo {title} {Quantum computing with parafermions},\
  }\href {https://doi.org/10.1103/PhysRevB.93.125105} {\bibfield  {journal}
  {\bibinfo  {journal} {Phys. Rev. B}\ }\textbf {\bibinfo {volume} {93}},\
  \bibinfo {pages} {125105} (\bibinfo {year} {2016})}\BibitemShut {NoStop}%
\bibitem [{\citenamefont {Cobanera}\ \emph {et~al.}(2016)\citenamefont
  {Cobanera}, \citenamefont {Ulrich},\ and\ \citenamefont
  {Hassler}}]{Hassler:ChangingAnyonicDegen:2016}%
  \BibitemOpen
  \bibfield  {author} {\bibinfo {author} {\bibfnamefont {E.}~\bibnamefont
  {Cobanera}}, \bibinfo {author} {\bibfnamefont {J.}~\bibnamefont {Ulrich}},\
  and\ \bibinfo {author} {\bibfnamefont {F.}~\bibnamefont {Hassler}},\
  }\bibfield  {title} {\bibinfo {title} {Changing anyonic ground degeneracy
  with engineered gauge fields},\ }\href
  {https://doi.org/10.1103/PhysRevB.94.125434} {\bibfield  {journal} {\bibinfo
  {journal} {Phys. Rev. B}\ }\textbf {\bibinfo {volume} {94}},\ \bibinfo
  {pages} {125434} (\bibinfo {year} {2016})}\BibitemShut {NoStop}%
\bibitem [{\citenamefont {Fradkin}\ and\ \citenamefont
  {Kadanoff}(1980)}]{Fradkin:Parafermions:1980}%
  \BibitemOpen
  \bibfield  {author} {\bibinfo {author} {\bibfnamefont {E.}~\bibnamefont
  {Fradkin}}\ and\ \bibinfo {author} {\bibfnamefont {L.}~\bibnamefont
  {Kadanoff}},\ }\bibfield  {title} {\bibinfo {title} {Disorder variables and
  para-fermions in two-dimensional statistical mechanics},\ }\href
  {https://doi.org/10.1016/0550-3213(80)90472-1} {\bibfield  {journal}
  {\bibinfo  {journal} {Nuclear Physics B}\ }\textbf {\bibinfo {volume}
  {170}},\ \bibinfo {pages} {1} (\bibinfo {year} {1980})}\BibitemShut {NoStop}%
\bibitem [{\citenamefont {Fendley}(2012)}]{Fendley:ParafermionicZeroMode:2012}%
  \BibitemOpen
  \bibfield  {author} {\bibinfo {author} {\bibfnamefont {P.}~\bibnamefont
  {Fendley}},\ }\bibfield  {title} {\bibinfo {title} {Parafermionic edge zero
  modes in z n -invariant spin chains},\ }\href
  {http://stacks.iop.org/1742-5468/2012/i=11/a=P11020} {\bibfield  {journal}
  {\bibinfo  {journal} {Journal of Statistical Mechanics: Theory and
  Experiment}\ }\textbf {\bibinfo {volume} {2012}},\ \bibinfo {pages} {P11020}
  (\bibinfo {year} {2012})}\BibitemShut {NoStop}%
\bibitem [{\citenamefont {Fleckenstein}\ \emph {et~al.}(2019)\citenamefont
  {Fleckenstein}, \citenamefont {Ziani},\ and\ \citenamefont
  {Trauzettel}}]{Fleckenstein:ParafermionHEdgeQSHI:2018}%
  \BibitemOpen
  \bibfield  {author} {\bibinfo {author} {\bibfnamefont {C.}~\bibnamefont
  {Fleckenstein}}, \bibinfo {author} {\bibfnamefont {N.~T.}\ \bibnamefont
  {Ziani}},\ and\ \bibinfo {author} {\bibfnamefont {B.}~\bibnamefont
  {Trauzettel}},\ }\bibfield  {title} {\bibinfo {title} {${\mathbb{z}}_{4}$
  parafermions in weakly interacting superconducting constrictions at the
  helical edge of quantum spin hall insulators},\ }\href
  {https://doi.org/10.1103/PhysRevLett.122.066801} {\bibfield  {journal}
  {\bibinfo  {journal} {Phys. Rev. Lett.}\ }\textbf {\bibinfo {volume} {122}},\
  \bibinfo {pages} {066801} (\bibinfo {year} {2019})}\BibitemShut {NoStop}%
\bibitem [{\citenamefont {Alavirad}\ \emph {et~al.}(2017)\citenamefont
  {Alavirad}, \citenamefont {Clarke}, \citenamefont {Nag},\ and\ \citenamefont
  {Sau}}]{Alavirad:Z3ParafermionWithoutAndreevFQH:2017}%
  \BibitemOpen
  \bibfield  {author} {\bibinfo {author} {\bibfnamefont {Y.}~\bibnamefont
  {Alavirad}}, \bibinfo {author} {\bibfnamefont {D.}~\bibnamefont {Clarke}},
  \bibinfo {author} {\bibfnamefont {A.}~\bibnamefont {Nag}},\ and\ \bibinfo
  {author} {\bibfnamefont {J.~D.}\ \bibnamefont {Sau}},\ }\bibfield  {title}
  {\bibinfo {title} {${\mathbb{z}}_{3}$ parafermionic zero modes without
  andreev backscattering from the $2/3$ fractional quantum hall state},\ }\href
  {https://doi.org/10.1103/PhysRevLett.119.217701} {\bibfield  {journal}
  {\bibinfo  {journal} {Phys. Rev. Lett.}\ }\textbf {\bibinfo {volume} {119}},\
  \bibinfo {pages} {217701} (\bibinfo {year} {2017})}\BibitemShut {NoStop}%
\bibitem [{\citenamefont {Vaezi}(2014)}]{Vaezi:SuperconductingPFQH:2014}%
  \BibitemOpen
  \bibfield  {author} {\bibinfo {author} {\bibfnamefont {A.}~\bibnamefont
  {Vaezi}},\ }\bibfield  {title} {\bibinfo {title} {Superconducting analogue of
  the parafermion fractional quantum hall states},\ }\href
  {https://doi.org/10.1103/PhysRevX.4.031009} {\bibfield  {journal} {\bibinfo
  {journal} {Phys. Rev. X}\ }\textbf {\bibinfo {volume} {4}},\ \bibinfo {pages}
  {031009} (\bibinfo {year} {2014})}\BibitemShut {NoStop}%
\bibitem [{\citenamefont {Vinkler-Aviv}\ \emph {et~al.}(2017)\citenamefont
  {Vinkler-Aviv}, \citenamefont {Brouwer},\ and\ \citenamefont {von
  Oppen}}]{vonOppen:Z4parafermionQSHJosephsonImp:2017}%
  \BibitemOpen
  \bibfield  {author} {\bibinfo {author} {\bibfnamefont {Y.}~\bibnamefont
  {Vinkler-Aviv}}, \bibinfo {author} {\bibfnamefont {P.~W.}\ \bibnamefont
  {Brouwer}},\ and\ \bibinfo {author} {\bibfnamefont {F.}~\bibnamefont {von
  Oppen}},\ }\bibfield  {title} {\bibinfo {title} {${\mathbb{z}}_{4}$
  parafermions in an interacting quantum spin hall josephson junction coupled
  to an impurity spin},\ }\href {https://doi.org/10.1103/PhysRevB.96.195421}
  {\bibfield  {journal} {\bibinfo  {journal} {Phys. Rev. B}\ }\textbf {\bibinfo
  {volume} {96}},\ \bibinfo {pages} {195421} (\bibinfo {year}
  {2017})}\BibitemShut {NoStop}%
\bibitem [{\citenamefont {Klinovaja}\ and\ \citenamefont
  {Loss}(2014{\natexlab{a}})}]{Klinovaja:InteractingNanowire:2014}%
  \BibitemOpen
  \bibfield  {author} {\bibinfo {author} {\bibfnamefont {J.}~\bibnamefont
  {Klinovaja}}\ and\ \bibinfo {author} {\bibfnamefont {D.}~\bibnamefont
  {Loss}},\ }\bibfield  {title} {\bibinfo {title} {Parafermions in an
  interacting nanowire bundle},\ }\href
  {https://doi.org/10.1103/PhysRevLett.112.246403} {\bibfield  {journal}
  {\bibinfo  {journal} {Phys. Rev. Lett.}\ }\textbf {\bibinfo {volume} {112}},\
  \bibinfo {pages} {246403} (\bibinfo {year} {2014}{\natexlab{a}})}\BibitemShut
  {NoStop}%
\bibitem [{\citenamefont {Read}\ and\ \citenamefont
  {Rezayi}(1999)}]{Read:ParafermionIncompressibleStates:1999}%
  \BibitemOpen
  \bibfield  {author} {\bibinfo {author} {\bibfnamefont {N.}~\bibnamefont
  {Read}}\ and\ \bibinfo {author} {\bibfnamefont {E.}~\bibnamefont {Rezayi}},\
  }\bibfield  {title} {\bibinfo {title} {Beyond paired quantum hall states:
  Parafermions and incompressible states in the first excited landau level},\
  }\href {https://doi.org/10.1103/PhysRevB.59.8084} {\bibfield  {journal}
  {\bibinfo  {journal} {Phys. Rev. B}\ }\textbf {\bibinfo {volume} {59}},\
  \bibinfo {pages} {8084} (\bibinfo {year} {1999})}\BibitemShut {NoStop}%
\bibitem [{\citenamefont {Jermyn}\ \emph {et~al.}(2014)\citenamefont {Jermyn},
  \citenamefont {Mong}, \citenamefont {Alicea},\ and\ \citenamefont
  {Fendley}}]{Fendley:StabilityParafermion:2014}%
  \BibitemOpen
  \bibfield  {author} {\bibinfo {author} {\bibfnamefont {A.~S.}\ \bibnamefont
  {Jermyn}}, \bibinfo {author} {\bibfnamefont {R.~S.~K.}\ \bibnamefont {Mong}},
  \bibinfo {author} {\bibfnamefont {J.}~\bibnamefont {Alicea}},\ and\ \bibinfo
  {author} {\bibfnamefont {P.}~\bibnamefont {Fendley}},\ }\bibfield  {title}
  {\bibinfo {title} {Stability of zero modes in parafermion chains},\ }\href
  {https://doi.org/10.1103/PhysRevB.90.165106} {\bibfield  {journal} {\bibinfo
  {journal} {Phys. Rev. B}\ }\textbf {\bibinfo {volume} {90}},\ \bibinfo
  {pages} {165106} (\bibinfo {year} {2014})}\BibitemShut {NoStop}%
\bibitem [{\citenamefont {Mong}\ \emph {et~al.}(2014)\citenamefont {Mong},
  \citenamefont {Clarke}, \citenamefont {Alicea}, \citenamefont {Lindner},
  \citenamefont {Fendley}, \citenamefont {Nayak}, \citenamefont {Oreg},
  \citenamefont {Stern}, \citenamefont {Berg}, \citenamefont {Shtengel},\ and\
  \citenamefont {Fisher}}]{Mong:UniversalTQC:2014}%
  \BibitemOpen
  \bibfield  {author} {\bibinfo {author} {\bibfnamefont {R.~S.~K.}\
  \bibnamefont {Mong}}, \bibinfo {author} {\bibfnamefont {D.~J.}\ \bibnamefont
  {Clarke}}, \bibinfo {author} {\bibfnamefont {J.}~\bibnamefont {Alicea}},
  \bibinfo {author} {\bibfnamefont {N.~H.}\ \bibnamefont {Lindner}}, \bibinfo
  {author} {\bibfnamefont {P.}~\bibnamefont {Fendley}}, \bibinfo {author}
  {\bibfnamefont {C.}~\bibnamefont {Nayak}}, \bibinfo {author} {\bibfnamefont
  {Y.}~\bibnamefont {Oreg}}, \bibinfo {author} {\bibfnamefont {A.}~\bibnamefont
  {Stern}}, \bibinfo {author} {\bibfnamefont {E.}~\bibnamefont {Berg}},
  \bibinfo {author} {\bibfnamefont {K.}~\bibnamefont {Shtengel}},\ and\
  \bibinfo {author} {\bibfnamefont {M.~P.~A.}\ \bibnamefont {Fisher}},\
  }\bibfield  {title} {\bibinfo {title} {Universal topological quantum
  computation from a superconductor-abelian quantum hall heterostructure},\
  }\href {https://doi.org/10.1103/PhysRevX.4.011036} {\bibfield  {journal}
  {\bibinfo  {journal} {Phys. Rev. X}\ }\textbf {\bibinfo {volume} {4}},\
  \bibinfo {pages} {011036} (\bibinfo {year} {2014})}\BibitemShut {NoStop}%
\bibitem [{\citenamefont {Clarke}\ \emph {et~al.}(2013)\citenamefont {Clarke},
  \citenamefont {Alicea},\ and\ \citenamefont
  {Shtengel}}]{Clarke:Non-AbelianAyonsFQH:2013}%
  \BibitemOpen
  \bibfield  {author} {\bibinfo {author} {\bibfnamefont {D.~J.}\ \bibnamefont
  {Clarke}}, \bibinfo {author} {\bibfnamefont {J.}~\bibnamefont {Alicea}},\
  and\ \bibinfo {author} {\bibfnamefont {K.}~\bibnamefont {Shtengel}},\
  }\bibfield  {title} {\bibinfo {title} {Exotic non-abelian anyons from
  conventional fractional quantum hall states},\ }\href
  {https://doi.org/10.1038/ncomms2340} {\bibfield  {journal} {\bibinfo
  {journal} {Nature Communications}\ }\textbf {\bibinfo {volume} {4}},\
  \bibinfo {pages} {1348 EP } (\bibinfo {year} {2013})},\ \bibinfo {note}
  {article}\BibitemShut {NoStop}%
\bibitem [{\citenamefont {Alicea}\ and\ \citenamefont
  {Fendley}(2016)}]{Alicea:TopologicalPhasesWithParafermions:2016}%
  \BibitemOpen
  \bibfield  {author} {\bibinfo {author} {\bibfnamefont {J.}~\bibnamefont
  {Alicea}}\ and\ \bibinfo {author} {\bibfnamefont {P.}~\bibnamefont
  {Fendley}},\ }\bibfield  {title} {\bibinfo {title} {Topological phases with
  parafermions: Theory and blueprints},\ }\href
  {https://doi.org/10.1146/annurev-conmatphys-031115-011336} {\bibfield
  {journal} {\bibinfo  {journal} {Annual Review of Condensed Matter Physics}\
  }\textbf {\bibinfo {volume} {7}},\ \bibinfo {pages} {119} (\bibinfo {year}
  {2016})}\BibitemShut {NoStop}%
\bibitem [{\citenamefont {Klinovaja}\ and\ \citenamefont
  {Loss}(2014{\natexlab{b}})}]{Klinovaja:TRIParafermionRashba:2014}%
  \BibitemOpen
  \bibfield  {author} {\bibinfo {author} {\bibfnamefont {J.}~\bibnamefont
  {Klinovaja}}\ and\ \bibinfo {author} {\bibfnamefont {D.}~\bibnamefont
  {Loss}},\ }\bibfield  {title} {\bibinfo {title} {Time-reversal invariant
  parafermions in interacting rashba nanowires},\ }\href
  {https://doi.org/10.1103/PhysRevB.90.045118} {\bibfield  {journal} {\bibinfo
  {journal} {Phys. Rev. B}\ }\textbf {\bibinfo {volume} {90}},\ \bibinfo
  {pages} {045118} (\bibinfo {year} {2014}{\natexlab{b}})}\BibitemShut
  {NoStop}%
\bibitem [{\citenamefont {Santos}\ and\ \citenamefont
  {Hughes}(2017)}]{Hughes:ParafermionicWiresCTstates:2017}%
  \BibitemOpen
  \bibfield  {author} {\bibinfo {author} {\bibfnamefont {L.~H.}\ \bibnamefont
  {Santos}}\ and\ \bibinfo {author} {\bibfnamefont {T.~L.}\ \bibnamefont
  {Hughes}},\ }\bibfield  {title} {\bibinfo {title} {Parafermionic wires at the
  interface of chiral topological states},\ }\href
  {https://doi.org/10.1103/PhysRevLett.118.136801} {\bibfield  {journal}
  {\bibinfo  {journal} {Phys. Rev. Lett.}\ }\textbf {\bibinfo {volume} {118}},\
  \bibinfo {pages} {136801} (\bibinfo {year} {2017})}\BibitemShut {NoStop}%
\bibitem [{\citenamefont {Klinovaja}\ \emph {et~al.}(2014)\citenamefont
  {Klinovaja}, \citenamefont {Yacoby},\ and\ \citenamefont
  {Loss}}]{Klinovaja:ParafermionsFTI:2014}%
  \BibitemOpen
  \bibfield  {author} {\bibinfo {author} {\bibfnamefont {J.}~\bibnamefont
  {Klinovaja}}, \bibinfo {author} {\bibfnamefont {A.}~\bibnamefont {Yacoby}},\
  and\ \bibinfo {author} {\bibfnamefont {D.}~\bibnamefont {Loss}},\ }\bibfield
  {title} {\bibinfo {title} {Kramers pairs of majorana fermions and
  parafermions in fractional topological insulators},\ }\href
  {https://doi.org/10.1103/PhysRevB.90.155447} {\bibfield  {journal} {\bibinfo
  {journal} {Phys. Rev. B}\ }\textbf {\bibinfo {volume} {90}},\ \bibinfo
  {pages} {155447} (\bibinfo {year} {2014})}\BibitemShut {NoStop}%
\bibitem [{\citenamefont {Kane}\ and\ \citenamefont
  {Zhang}(2015)}]{Kane:Poorman:2015}%
  \BibitemOpen
  \bibfield  {author} {\bibinfo {author} {\bibfnamefont {C.~L.}\ \bibnamefont
  {Kane}}\ and\ \bibinfo {author} {\bibfnamefont {F.}~\bibnamefont {Zhang}},\
  }\bibfield  {title} {\bibinfo {title} {The time reversal invariant fractional
  josephson effect},\ }\href
  {https://doi.org/10.1088/0031-8949/2015/t164/014011} {\bibfield  {journal}
  {\bibinfo  {journal} {Physica Scripta}\ }\textbf {\bibinfo {volume} {T164}},\
  \bibinfo {pages} {014011} (\bibinfo {year} {2015})}\BibitemShut {NoStop}%
\bibitem [{\citenamefont {Rossini}\ \emph {et~al.}(2019)\citenamefont
  {Rossini}, \citenamefont {Carrega}, \citenamefont {Calvanese~Strinati},\ and\
  \citenamefont {Mazza}}]{Mazza:AnyonicTBofparafermions:2019}%
  \BibitemOpen
  \bibfield  {author} {\bibinfo {author} {\bibfnamefont {D.}~\bibnamefont
  {Rossini}}, \bibinfo {author} {\bibfnamefont {M.}~\bibnamefont {Carrega}},
  \bibinfo {author} {\bibfnamefont {M.}~\bibnamefont {Calvanese~Strinati}},\
  and\ \bibinfo {author} {\bibfnamefont {L.}~\bibnamefont {Mazza}},\ }\bibfield
   {title} {\bibinfo {title} {Anyonic tight-binding models of parafermions and
  of fractionalized fermions},\ }\href
  {https://doi.org/10.1103/PhysRevB.99.085113} {\bibfield  {journal} {\bibinfo
  {journal} {Phys. Rev. B}\ }\textbf {\bibinfo {volume} {99}},\ \bibinfo
  {pages} {085113} (\bibinfo {year} {2019})}\BibitemShut {NoStop}%
\bibitem [{\citenamefont {Mazza}\ \emph {et~al.}(2018)\citenamefont {Mazza},
  \citenamefont {Iemini}, \citenamefont {Dalmonte},\ and\ \citenamefont
  {Mora}}]{Mazza:NonTopological:2018}%
  \BibitemOpen
  \bibfield  {author} {\bibinfo {author} {\bibfnamefont {L.}~\bibnamefont
  {Mazza}}, \bibinfo {author} {\bibfnamefont {F.}~\bibnamefont {Iemini}},
  \bibinfo {author} {\bibfnamefont {M.}~\bibnamefont {Dalmonte}},\ and\
  \bibinfo {author} {\bibfnamefont {C.}~\bibnamefont {Mora}},\ }\bibfield
  {title} {\bibinfo {title} {Nontopological parafermions in a one-dimensional
  fermionic model with even multiplet pairing},\ }\href
  {https://doi.org/10.1103/PhysRevB.98.201109} {\bibfield  {journal} {\bibinfo
  {journal} {Phys. Rev. B}\ }\textbf {\bibinfo {volume} {98}},\ \bibinfo
  {pages} {201109(R)} (\bibinfo {year} {2018})}\BibitemShut {NoStop}%
\bibitem [{\citenamefont {Chew}\ \emph {et~al.}(2018)\citenamefont {Chew},
  \citenamefont {Mross},\ and\ \citenamefont
  {Alicea}}]{Alicea:FermionezParafermionsSEMajorana:2018}%
  \BibitemOpen
  \bibfield  {author} {\bibinfo {author} {\bibfnamefont {A.}~\bibnamefont
  {Chew}}, \bibinfo {author} {\bibfnamefont {D.~F.}\ \bibnamefont {Mross}},\
  and\ \bibinfo {author} {\bibfnamefont {J.}~\bibnamefont {Alicea}},\
  }\bibfield  {title} {\bibinfo {title} {Fermionized parafermions and
  symmetry-enriched majorana modes},\ }\href
  {https://doi.org/10.1103/PhysRevB.98.085143} {\bibfield  {journal} {\bibinfo
  {journal} {Phys. Rev. B}\ }\textbf {\bibinfo {volume} {98}},\ \bibinfo
  {pages} {085143} (\bibinfo {year} {2018})}\BibitemShut {NoStop}%
\bibitem [{\citenamefont {Alexandradinata}\ \emph {et~al.}(2016)\citenamefont
  {Alexandradinata}, \citenamefont {Regnault}, \citenamefont {Fang},
  \citenamefont {Gilbert},\ and\ \citenamefont
  {Bernevig}}]{Bernevig:ParafermionPhasesSbreakingTO:2016}%
  \BibitemOpen
  \bibfield  {author} {\bibinfo {author} {\bibfnamefont {A.}~\bibnamefont
  {Alexandradinata}}, \bibinfo {author} {\bibfnamefont {N.}~\bibnamefont
  {Regnault}}, \bibinfo {author} {\bibfnamefont {C.}~\bibnamefont {Fang}},
  \bibinfo {author} {\bibfnamefont {M.~J.}\ \bibnamefont {Gilbert}},\ and\
  \bibinfo {author} {\bibfnamefont {B.~A.}\ \bibnamefont {Bernevig}},\
  }\bibfield  {title} {\bibinfo {title} {Parafermionic phases with symmetry
  breaking and topological order},\ }\href
  {https://doi.org/10.1103/PhysRevB.94.125103} {\bibfield  {journal} {\bibinfo
  {journal} {Phys. Rev. B}\ }\textbf {\bibinfo {volume} {94}},\ \bibinfo
  {pages} {125103} (\bibinfo {year} {2016})}\BibitemShut {NoStop}%
\bibitem [{\citenamefont {Calzona}\ \emph {et~al.}(2018)\citenamefont
  {Calzona}, \citenamefont {Meng}, \citenamefont {Sassetti},\ and\
  \citenamefont {Schmidt}}]{Calzona:Z4FermionLattice:2018}%
  \BibitemOpen
  \bibfield  {author} {\bibinfo {author} {\bibfnamefont {A.}~\bibnamefont
  {Calzona}}, \bibinfo {author} {\bibfnamefont {T.}~\bibnamefont {Meng}},
  \bibinfo {author} {\bibfnamefont {M.}~\bibnamefont {Sassetti}},\ and\
  \bibinfo {author} {\bibfnamefont {T.~L.}\ \bibnamefont {Schmidt}},\
  }\bibfield  {title} {\bibinfo {title} {${\mathbb{z}}_{4}$ parafermions in
  one-dimensional fermionic lattices},\ }\href
  {https://doi.org/10.1103/PhysRevB.98.201110} {\bibfield  {journal} {\bibinfo
  {journal} {Phys. Rev. B}\ }\textbf {\bibinfo {volume} {98}},\ \bibinfo
  {pages} {201110(R)} (\bibinfo {year} {2018})}\BibitemShut {NoStop}%
\bibitem [{\citenamefont {Vernek}\ \emph {et~al.}(2014)\citenamefont {Vernek},
  \citenamefont {Penteado}, \citenamefont {Seridonio},\ and\ \citenamefont
  {Egues}}]{Vernek:MajoarnaLeakage:2014}%
  \BibitemOpen
  \bibfield  {author} {\bibinfo {author} {\bibfnamefont {E.}~\bibnamefont
  {Vernek}}, \bibinfo {author} {\bibfnamefont {P.~H.}\ \bibnamefont
  {Penteado}}, \bibinfo {author} {\bibfnamefont {A.~C.}\ \bibnamefont
  {Seridonio}},\ and\ \bibinfo {author} {\bibfnamefont {J.~C.}\ \bibnamefont
  {Egues}},\ }\bibfield  {title} {\bibinfo {title} {Subtle leakage of a
  majorana mode into a quantum dot},\ }\href
  {https://doi.org/10.1103/PhysRevB.89.165314} {\bibfield  {journal} {\bibinfo
  {journal} {Phys. Rev. B}\ }\textbf {\bibinfo {volume} {89}},\ \bibinfo
  {pages} {165314} (\bibinfo {year} {2014})}\BibitemShut {NoStop}%
\bibitem [{\citenamefont {Ruiz-Tijerina}\ \emph {et~al.}(2015)\citenamefont
  {Ruiz-Tijerina}, \citenamefont {Vernek}, \citenamefont {Dias~da Silva},\ and\
  \citenamefont {Egues}}]{David:InteractionsMajoranaQD:2015}%
  \BibitemOpen
  \bibfield  {author} {\bibinfo {author} {\bibfnamefont {D.~A.}\ \bibnamefont
  {Ruiz-Tijerina}}, \bibinfo {author} {\bibfnamefont {E.}~\bibnamefont
  {Vernek}}, \bibinfo {author} {\bibfnamefont {L.~G. G.~V.}\ \bibnamefont
  {Dias~da Silva}},\ and\ \bibinfo {author} {\bibfnamefont {J.~C.}\
  \bibnamefont {Egues}},\ }\bibfield  {title} {\bibinfo {title} {Interaction
  effects on a majorana zero mode leaking into a quantum dot},\ }\href
  {https://doi.org/10.1103/PhysRevB.91.115435} {\bibfield  {journal} {\bibinfo
  {journal} {Phys. Rev. B}\ }\textbf {\bibinfo {volume} {91}},\ \bibinfo
  {pages} {115435} (\bibinfo {year} {2015})}\BibitemShut {NoStop}%
\bibitem [{\citenamefont {Teixeira}\ and\ \citenamefont {Dias~da
  Silva}(2021)}]{Teixeira:QD:2021}%
  \BibitemOpen
  \bibfield  {author} {\bibinfo {author} {\bibfnamefont {R.~L. R.~C.}\
  \bibnamefont {Teixeira}}\ and\ \bibinfo {author} {\bibfnamefont {L.~G.
  G.~V.}\ \bibnamefont {Dias~da Silva}},\ }\bibfield  {title} {\bibinfo {title}
  {Quantum dots as parafermion detectors},\ }\href
  {https://doi.org/10.1103/PhysRevResearch.3.033014} {\bibfield  {journal}
  {\bibinfo  {journal} {Phys. Rev. Research}\ }\textbf {\bibinfo {volume}
  {3}},\ \bibinfo {pages} {033014} (\bibinfo {year} {2021})}\BibitemShut
  {NoStop}%
\bibitem [{\citenamefont {Oreg}\ \emph {et~al.}(2010)\citenamefont {Oreg},
  \citenamefont {Refael},\ and\ \citenamefont {von
  Oppen}}]{vonOppen:HelicalLiquidsMajoranaQW:2010}%
  \BibitemOpen
  \bibfield  {author} {\bibinfo {author} {\bibfnamefont {Y.}~\bibnamefont
  {Oreg}}, \bibinfo {author} {\bibfnamefont {G.}~\bibnamefont {Refael}},\ and\
  \bibinfo {author} {\bibfnamefont {F.}~\bibnamefont {von Oppen}},\ }\bibfield
  {title} {\bibinfo {title} {Helical liquids and majorana bound states in
  quantum wires},\ }\href {https://doi.org/10.1103/PhysRevLett.105.177002}
  {\bibfield  {journal} {\bibinfo  {journal} {Phys. Rev. Lett.}\ }\textbf
  {\bibinfo {volume} {105}},\ \bibinfo {pages} {177002} (\bibinfo {year}
  {2010})}\BibitemShut {NoStop}%
\bibitem [{\citenamefont {Lutchyn}\ \emph {et~al.}(2010)\citenamefont
  {Lutchyn}, \citenamefont {Sau},\ and\ \citenamefont
  {Das~Sarma}}]{DasSarma:MajoranaPhaseSemiconductor:2010}%
  \BibitemOpen
  \bibfield  {author} {\bibinfo {author} {\bibfnamefont {R.~M.}\ \bibnamefont
  {Lutchyn}}, \bibinfo {author} {\bibfnamefont {J.~D.}\ \bibnamefont {Sau}},\
  and\ \bibinfo {author} {\bibfnamefont {S.}~\bibnamefont {Das~Sarma}},\
  }\bibfield  {title} {\bibinfo {title} {Majorana fermions and a topological
  phase transition in semiconductor-superconductor heterostructures},\ }\href
  {https://doi.org/10.1103/PhysRevLett.105.077001} {\bibfield  {journal}
  {\bibinfo  {journal} {Phys. Rev. Lett.}\ }\textbf {\bibinfo {volume} {105}},\
  \bibinfo {pages} {077001} (\bibinfo {year} {2010})}\BibitemShut {NoStop}%
\bibitem [{\citenamefont {Alicea}(2012)}]{Alicea:Reports:2012}%
  \BibitemOpen
  \bibfield  {author} {\bibinfo {author} {\bibfnamefont {J.}~\bibnamefont
  {Alicea}},\ }\bibfield  {title} {\bibinfo {title} {New directions in the
  pursuit of majorana fermions in solid state systems},\ }\href
  {http://stacks.iop.org/0034-4885/75/i=7/a=076501} {\bibfield  {journal}
  {\bibinfo  {journal} {Rep. Prog. Phys.}\ }\textbf {\bibinfo {volume} {75}},\
  \bibinfo {pages} {076501} (\bibinfo {year} {2012})}\BibitemShut {NoStop}%
\bibitem [{\citenamefont {Schollw\"ock}(2005)}]{Schollwock:DMRG:2005}%
  \BibitemOpen
  \bibfield  {author} {\bibinfo {author} {\bibfnamefont {U.}~\bibnamefont
  {Schollw\"ock}},\ }\bibfield  {title} {\bibinfo {title} {The density-matrix
  renormalization group},\ }\href {https://doi.org/10.1103/RevModPhys.77.259}
  {\bibfield  {journal} {\bibinfo  {journal} {Rev. Mod. Phys.}\ }\textbf
  {\bibinfo {volume} {77}},\ \bibinfo {pages} {259} (\bibinfo {year}
  {2005})}\BibitemShut {NoStop}%
\bibitem [{\citenamefont {Schollw\"ock}(2011)}]{Schollwock:DMRG-MPS:2011}%
  \BibitemOpen
  \bibfield  {author} {\bibinfo {author} {\bibfnamefont {U.}~\bibnamefont
  {Schollw\"ock}},\ }\bibfield  {title} {\bibinfo {title} {The density-matrix
  renormalization group in the age of matrix product states},\ }\href
  {https://doi.org/https://doi.org/10.1016/j.aop.2010.09.012} {\bibfield
  {journal} {\bibinfo  {journal} {Annals of Physics}\ }\textbf {\bibinfo
  {volume} {326}},\ \bibinfo {pages} {96 } (\bibinfo {year} {2011})},\ \bibinfo
  {note} {january 2011 Special Issue}\BibitemShut {NoStop}%
\bibitem [{\citenamefont {Fishman}\ \emph {et~al.}(2020)\citenamefont
  {Fishman}, \citenamefont {White},\ and\ \citenamefont
  {Stoudenmire}}]{ITensor}%
  \BibitemOpen
  \bibfield  {author} {\bibinfo {author} {\bibfnamefont {M.}~\bibnamefont
  {Fishman}}, \bibinfo {author} {\bibfnamefont {S.~R.}\ \bibnamefont {White}},\
  and\ \bibinfo {author} {\bibfnamefont {E.~M.}\ \bibnamefont {Stoudenmire}},\
  }\href@noop {} {\bibinfo {title} {The \mbox{ITensor} software library for
  tensor network calculations}} (\bibinfo {year} {2020}),\ \Eprint
  {https://arxiv.org/abs/2007.14822} {arXiv:2007.14822} \BibitemShut {NoStop}%
\bibitem [{\citenamefont {Batista}\ and\ \citenamefont
  {Ortiz}(2001)}]{Batista:GeneralizedJW:2001}%
  \BibitemOpen
  \bibfield  {author} {\bibinfo {author} {\bibfnamefont {C.~D.}\ \bibnamefont
  {Batista}}\ and\ \bibinfo {author} {\bibfnamefont {G.}~\bibnamefont
  {Ortiz}},\ }\bibfield  {title} {\bibinfo {title} {Generalized jordan-wigner
  transformations},\ }\href {https://doi.org/10.1103/PhysRevLett.86.1082}
  {\bibfield  {journal} {\bibinfo  {journal} {Phys. Rev. Lett.}\ }\textbf
  {\bibinfo {volume} {86}},\ \bibinfo {pages} {1082} (\bibinfo {year}
  {2001})}\BibitemShut {NoStop}%
\bibitem [{\citenamefont {Zhuang}\ \emph {et~al.}(2015)\citenamefont {Zhuang},
  \citenamefont {Changlani}, \citenamefont {Tubman},\ and\ \citenamefont
  {Hughes}}]{Zhuang:PhaseDiagramZ3Parafermion:2015}%
  \BibitemOpen
  \bibfield  {author} {\bibinfo {author} {\bibfnamefont {Y.}~\bibnamefont
  {Zhuang}}, \bibinfo {author} {\bibfnamefont {H.~J.}\ \bibnamefont
  {Changlani}}, \bibinfo {author} {\bibfnamefont {N.~M.}\ \bibnamefont
  {Tubman}},\ and\ \bibinfo {author} {\bibfnamefont {T.~L.}\ \bibnamefont
  {Hughes}},\ }\bibfield  {title} {\bibinfo {title} {Phase diagram of the
  ${Z}_{3}$ parafermionic chain with chiral interactions},\ }\href
  {https://doi.org/10.1103/PhysRevB.92.035154} {\bibfield  {journal} {\bibinfo
  {journal} {Phys. Rev. B}\ }\textbf {\bibinfo {volume} {92}},\ \bibinfo
  {pages} {035154} (\bibinfo {year} {2015})}\BibitemShut {NoStop}%
\bibitem [{\citenamefont {Iemini}\ \emph {et~al.}(2017)\citenamefont {Iemini},
  \citenamefont {Mora},\ and\ \citenamefont
  {Mazza}}]{Mazza:ParafermionGS:2017}%
  \BibitemOpen
  \bibfield  {author} {\bibinfo {author} {\bibfnamefont {F.}~\bibnamefont
  {Iemini}}, \bibinfo {author} {\bibfnamefont {C.}~\bibnamefont {Mora}},\ and\
  \bibinfo {author} {\bibfnamefont {L.}~\bibnamefont {Mazza}},\ }\bibfield
  {title} {\bibinfo {title} {Topological phases of parafermions: A model with
  exactly solvable ground states},\ }\href
  {https://doi.org/10.1103/PhysRevLett.118.170402} {\bibfield  {journal}
  {\bibinfo  {journal} {Phys. Rev. Lett.}\ }\textbf {\bibinfo {volume} {118}},\
  \bibinfo {pages} {170402} (\bibinfo {year} {2017})}\BibitemShut {NoStop}%
\bibitem [{\citenamefont {Gong}\ \emph {et~al.}(2016)\citenamefont {Gong},
  \citenamefont {Maghrebi}, \citenamefont {Hu}, \citenamefont {Wall},
  \citenamefont {Foss-Feig},\ and\ \citenamefont
  {Gorshkov}}]{Gong:Phys.Rev.B:041102:2016}%
  \BibitemOpen
  \bibfield  {author} {\bibinfo {author} {\bibfnamefont {Z.-X.}\ \bibnamefont
  {Gong}}, \bibinfo {author} {\bibfnamefont {M.~F.}\ \bibnamefont {Maghrebi}},
  \bibinfo {author} {\bibfnamefont {A.}~\bibnamefont {Hu}}, \bibinfo {author}
  {\bibfnamefont {M.~L.}\ \bibnamefont {Wall}}, \bibinfo {author}
  {\bibfnamefont {M.}~\bibnamefont {Foss-Feig}},\ and\ \bibinfo {author}
  {\bibfnamefont {A.~V.}\ \bibnamefont {Gorshkov}},\ }\bibfield  {title}
  {\bibinfo {title} {Topological phases with long-range interactions},\ }\href
  {https://doi.org/10.1103/PhysRevB.93.041102} {\bibfield  {journal} {\bibinfo
  {journal} {Phys. Rev. B}\ }\textbf {\bibinfo {volume} {93}},\ \bibinfo
  {pages} {041102(R)} (\bibinfo {year} {2016})}\BibitemShut {NoStop}%
\bibitem [{\citenamefont {Yu}\ \emph {et~al.}(2020)\citenamefont {Yu},
  \citenamefont {Cheng},\ and\ \citenamefont
  {Sacramento}}]{Yu:Phys.Rev.B:245131:2020}%
  \BibitemOpen
  \bibfield  {author} {\bibinfo {author} {\bibfnamefont {W.~C.}\ \bibnamefont
  {Yu}}, \bibinfo {author} {\bibfnamefont {C.}~\bibnamefont {Cheng}},\ and\
  \bibinfo {author} {\bibfnamefont {P.~D.}\ \bibnamefont {Sacramento}},\
  }\bibfield  {title} {\bibinfo {title} {Energy bonds as correlators for
  long-range symmetry-protected topological models and models with long-range
  topological order},\ }\href {https://doi.org/10.1103/PhysRevB.101.245131}
  {\bibfield  {journal} {\bibinfo  {journal} {Phys. Rev. B}\ }\textbf {\bibinfo
  {volume} {101}},\ \bibinfo {pages} {245131} (\bibinfo {year}
  {2020})}\BibitemShut {NoStop}%
\bibitem [{\citenamefont {White}(1993)}]{White:Phys.Rev.B:10345:1993}%
  \BibitemOpen
  \bibfield  {author} {\bibinfo {author} {\bibfnamefont {S.~R.}\ \bibnamefont
  {White}},\ }\bibfield  {title} {\bibinfo {title} {Density-matrix algorithms
  for quantum renormalization groups},\ }\href@noop {} {\bibfield  {journal}
  {\bibinfo  {journal} {Phys. Rev. B}\ }\textbf {\bibinfo {volume} {48}},\
  \bibinfo {pages} {10345} (\bibinfo {year} {1993})}\BibitemShut {NoStop}%
\bibitem [{Note1()}]{Note1}%
  \BibitemOpen
  \bibinfo {note} {Around $W_6\protect \!=\protect \!2t$ we used steps of
  $0.05t$ to plot figure \ref {fig:phases}(a).}\BibitemShut {Stop}%
\bibitem [{Note2()}]{Note2}%
  \BibitemOpen
  \bibinfo {note} {To compute the entanglement entropy, we perform DMRG
  calculations fully preserving the $\protect \mathbb {Z}_3$ symmetry. The
  entanglement entropy is then calculated at the system's center
  bond.}\BibitemShut {Stop}%
\bibitem [{Note3()}]{Note3}%
  \BibitemOpen
  \bibinfo {note} {The exponential decay happens as long as $\mu $ is such that
  does not break the $\protect \mathbb {Z}_3$ symmetry.}\BibitemShut {Stop}%
\bibitem [{\citenamefont {Das~Sarma}\ \emph {et~al.}(2012)\citenamefont
  {Das~Sarma}, \citenamefont {Sau},\ and\ \citenamefont
  {Stanescu}}]{DasSarma:SmokingGun:2012}%
  \BibitemOpen
  \bibfield  {author} {\bibinfo {author} {\bibfnamefont {S.}~\bibnamefont
  {Das~Sarma}}, \bibinfo {author} {\bibfnamefont {J.~D.}\ \bibnamefont {Sau}},\
  and\ \bibinfo {author} {\bibfnamefont {T.~D.}\ \bibnamefont {Stanescu}},\
  }\bibfield  {title} {\bibinfo {title} {Splitting of the zero-bias conductance
  peak as smoking gun evidence for the existence of the majorana mode in a
  superconductor-semiconductor nanowire},\ }\href
  {https://doi.org/10.1103/PhysRevB.86.220506} {\bibfield  {journal} {\bibinfo
  {journal} {Phys. Rev. B}\ }\textbf {\bibinfo {volume} {86}},\ \bibinfo
  {pages} {220506(R)} (\bibinfo {year} {2012})}\BibitemShut {NoStop}%
\bibitem [{\citenamefont {Li}\ \emph {et~al.}(2013)\citenamefont {Li},
  \citenamefont {Weichselbaum},\ and\ \citenamefont
  {vonDelft}}]{Li:Phys.Rev.B:245121:2013}%
  \BibitemOpen
  \bibfield  {author} {\bibinfo {author} {\bibfnamefont {W.}~\bibnamefont
  {Li}}, \bibinfo {author} {\bibfnamefont {A.}~\bibnamefont {Weichselbaum}},\
  and\ \bibinfo {author} {\bibfnamefont {J.}~\bibnamefont {vonDelft}},\
  }\bibfield  {title} {\bibinfo {title} {Identifying symmetry-protected
  topological order by entanglement entropy},\ }\href
  {https://doi.org/10.1103/PhysRevB.88.245121} {\bibfield  {journal} {\bibinfo
  {journal} {Phys. Rev. B}\ }\textbf {\bibinfo {volume} {88}},\ \bibinfo
  {pages} {245121} (\bibinfo {year} {2013})}\BibitemShut {NoStop}%
\bibitem [{\citenamefont {Cobanera}\ and\ \citenamefont
  {Ortiz}(2014)}]{Cobanera:FockParafermion:2014}%
  \BibitemOpen
  \bibfield  {author} {\bibinfo {author} {\bibfnamefont {E.}~\bibnamefont
  {Cobanera}}\ and\ \bibinfo {author} {\bibfnamefont {G.}~\bibnamefont
  {Ortiz}},\ }\bibfield  {title} {\bibinfo {title} {Fock parafermions and
  self-dual representations of the braid group},\ }\href
  {https://doi.org/10.1103/PhysRevA.89.012328} {\bibfield  {journal} {\bibinfo
  {journal} {Phys. Rev. A}\ }\textbf {\bibinfo {volume} {89}},\ \bibinfo
  {pages} {012328} (\bibinfo {year} {2014})}\BibitemShut {NoStop}%
\bibitem [{Note4()}]{Note4}%
  \BibitemOpen
  \bibinfo {note} {The problem arises in the $\protect \mathbb {Z}_3$ case
  mainly because there is no single fermionic operator connecting the up and
  down states. By contrast, in the case of $\protect \mathbb {Z}_4$
  parafermions, two sequential FPF numbers are connected by a single fermionic
  operator, eliminating the problem.}\BibitemShut {Stop}%
\bibitem [{Note5()}]{Note5}%
  \BibitemOpen
  \bibinfo {note} {Another option is to consider $d_j \protect \bar {\omega
  }^{N_j/2}$ which will also have a symmetric zero-energy spectral function.
  The biggest difference is a scaling factor and should not affect how the edge
  states are localized.}\BibitemShut {Stop}%
\end{thebibliography}

%

\end{document}